\documentclass[floats,floatfix,twocolumn,prx]{revtex4-2}
\usepackage{graphics}
\usepackage{amsmath}
\usepackage{graphicx}
\usepackage{amsfonts}
\usepackage{amssymb}
\usepackage{eurosym}
\usepackage{graphics}
\usepackage{epsfig}
\usepackage{latexsym}
\usepackage{theorem}
\usepackage{bm}
\usepackage{ulem}
\usepackage{color}
\usepackage{graphicx}
\usepackage{placeins}
\usepackage{tikz}
\usetikzlibrary{positioning,arrows.meta,fit,calc,backgrounds}
\usepackage{booktabs}
\usepackage{array}[=2016-10-06]
\usepackage{tabularx}
\usepackage{url}

\makeatletter
\long\def\@makecaption#1#2{%
  \par
  \vskip\abovecaptionskip
  \begingroup
   \small\rmfamily
   \begingroup
    \samepage
    \flushing
    \let\footnote\@footnotemark@gobble
    \@make@capt@title{#1}{#2}\par
   \endgroup
  \endgroup
  \vskip\belowcaptionskip
}%
\makeatother

\newcommand{\IP}[1]{{{\textcolor{black}{#1}}}}

\begin{document}

\title{\IP{Post-Quantum Cryptography in Data Centres: \\Fat-tree and Jellyfish Case Studies}}

\author{Ignazio Pedone}
\affiliation{nodeQ Limited, 71-75 Shelton Street, Covent Garden, London, WC2H 9JQ, United Kingdom}
\author{Stefano Pirandola}
\affiliation{Department of Computer Science, University of York, Deramore Lane, Heslington, York, YO10 5GH, United Kingdom}
\email{Corresponding author: stefano.pirandola@york.ac.uk}

\begin{abstract}
Migrating widely used security protocols to post-quantum cryptography can increase the processing and network resources required for secure connection establishment. Benchmarks of individual operations and endpoint connections do not establish how these costs affect networks where many connections share processing and link capacity. To examine these effects at network scale, we benchmark post-quantum TLS~1.3 handshake traffic on hierarchical fat-tree and random regular Jellyfish networks. These contrasting structures are evaluated under independent shortest-path and coordinated fractional routing as the number of connections increases. The traffic profiles use NIST-standardised ML-KEM and ML-DSA, with Classic McEliece providing a modelled stress case for transmitting large public keys during the handshake. Two main findings emerge. First, the benchmark establishes how the cryptographic profile affects demand coverage, network stress, and routing efficiency as load increases. The more demanding profiles achieve lower coverage and routing efficiency, with substantial unserved demand even where average link utilisation remains low. Second, coordinated routing yields larger and more sustained coverage gains on the tested Jellyfish network than on the fat-tree, indicating the potential of Jellyfish combined with fractional routing to mitigate the network impact of demanding post-quantum profiles. Together, these findings highlight the value of accounting for cryptographic processing and communication costs in routing and bandwidth allocation, alongside network capacity and endpoint placement.
\end{abstract}

\maketitle
\raggedbottom

\section{Introduction}

Large-scale communication systems are entering a transition period in which classical public-key cryptography must be replaced or hybridised with post-quantum alternatives. Shor's quantum algorithm~\cite{Shor1997} shows that integer factorisation and discrete logarithms can be solved efficiently on a sufficiently capable quantum computer~\cite{Shor}; consequently, widely deployed key-establishment and signature mechanisms based on RSA~\cite{RSA1978}, finite-field Diffie--Hellman~\cite{DiffieHellman1976}, elliptic-curve Diffie--Hellman~\cite{Miller1986}, and elliptic-curve signatures~\cite{Koblitz1987,FIPS186-5} are not long-term quantum-safe. NIST has consequently proposed a timetable for migration to post-quantum cryptography in its draft transition guidance~\cite{NISTIR8547}. The post-quantum migration is therefore not only an algorithm-selection problem, but also a systems problem: new cryptographic primitives must be deployed inside networks that already operate close to latency, throughput, routing-state, and congestion constraints.

The recent NIST PQC standards provide concrete primitives for this migration. ML-KEM is standardised for key establishment in FIPS~203~\cite{FIPS203}, ML-DSA for digital signatures in FIPS~204~\cite{FIPS204}, and SLH-DSA for stateless hash-based signatures in FIPS~205~\cite{FIPS205}. These standards make deployment technically possible, but they do not by themselves determine the network-level consequences of deployment. PQC mechanisms may differ from pre-quantum mechanisms in public-key size, ciphertext or signature size, handshake structure, and computational cost, and measurements on TLS and SSH testbeds have shown that these differences can translate into lower handshake rates and larger handshake payloads~\cite{Sikeridis2020,Sosnowski2023}. In a data centre, these differences can affect endpoint processing time, service placement, network congestion, and the amount of secure traffic that can be admitted under a given routing policy.

Routing and traffic engineering are therefore central to PQC deployment. Data-centre studies have shown that topology and routing choices strongly affect throughput, congestion, and the ability to exploit path diversity. Fat-tree and Clos-derived designs were introduced to provide high bisection bandwidth from commodity switches~\cite{AlFares}, while systems such as Hedera and MicroTE showed that traffic-aware scheduling and fine-grained traffic engineering can substantially improve data-centre performance under realistic traffic variation~\cite{Hedera,MicroTE}. Randomised topologies such as Jellyfish provide a complementary design point, using random regular connectivity to obtain flexible expansion and high path diversity~\cite{Jellyfish}. These results suggest that PQC overhead should not be evaluated only at the level of isolated cryptographic operations; it should also be evaluated through the routing and traffic-management layer that determines which secure flows can actually be carried.

Work on PQC deployment includes specifications of cryptographic primitives and their key, ciphertext, and signature sizes~\cite{FIPS203,FIPS204,FIPS205,ClassicMcEliece}, as well as measurements of handshake rates and payloads in PQC-enabled TLS and SSH testbeds~\cite{Sikeridis2020,Sosnowski2023}. The throughput model of Ref.~\cite{PQCmodel} connects endpoint processing and communication costs to sustainable handshake rates and extends the analysis to network resource allocation among multiple endpoints. Building on that model, the present study examines how PQC traffic classes and routing policies interact with different data-centre network structures as load increases.

In this paper, we combine the PQC endpoint and network throughput model of Ref.~\cite{PQCmodel} with multiple-unicast routing optimisation to study PQC traffic on fat-tree and Jellyfish networks. The model accounts for the processing operations and bytes transmitted during a TLS~1.3 handshake. It first determines the rates that the endpoints can sustain and converts them into bandwidth demands. After routing, it converts the allocated bandwidth back into served handshake rates. This allows the processing and communication costs of each cryptographic profile to be evaluated together with the capacity of the network.

We examine how demand coverage and the total served handshake rate change as the number of connections increases, how these trends differ between light and heavy cryptographic traffic, and how much additional demand coordinated routing can serve. The fat-tree provides a hierarchical fabric, while Jellyfish provides a random regular fabric with different route alternatives. Comparing routing policies on each network shows how effectively they use the available capacity. Comparing the resulting gains across the two deployments indicates how routing effectiveness depends on network structure and provisioning.

The study makes three contributions:
\begin{itemize}
    \item a common benchmark workflow that accounts for endpoint processing, converts PQC handshake requests into bandwidth demands, and evaluates routing on structured and random networks;
    \item a comparison of light ML-KEM/ML-DSA traffic and heavy Classic McEliece traffic as load increases, using OSPF-like, fractional $1$-path, and fractional $5$-path routing;
    \item an analysis of routing gains and possible capacity constraints using demand coverage, total served handshake rate, network stress, routing efficiency, and security composition. Policy comparisons identify the routing contribution, while achieved coverage bounds the loss from endpoint processing limits.
\end{itemize}

Two principal findings emerge. The first is the substantial difference between the traffic classes. The light class achieves almost complete coverage under coordinated routing. The heavy class requests about seven times more bandwidth per connection despite a tenfold lower handshake rate, and leaves substantial demand unserved from the lowest tested load. These results show why both the cryptographic profile and the requested rate matter when provisioning a network.

The second finding is that coordinated routing delivers substantial gains whose size and persistence differ between the two network structures. For heavy traffic, fractional $5$-path routing gains up to $24$ percentage points of coverage over OSPF-like routing for fat-tree east--west traffic and up to $30$ points on Jellyfish. On Jellyfish, it recovers up to about $70\%$ of the demand left unserved by the baseline, compared with less than $40\%$ for fat-tree east--west traffic. The Jellyfish gain remains substantial as load grows, while the fat-tree gain narrows. Fractional $1$-path routing captures most of these gains, showing that coordinated path selection and bandwidth allocation are effective even with one retained path per connection. The smaller gains for fat-tree north--south traffic depend more on retaining additional paths. These comparisons establish the benefit on the tested deployments; differences between them reflect capacity and endpoint placement as well as topology.

The remainder of the paper is organised as follows. Section~\ref{sec:topologies} introduces the network model and the fat-tree and Jellyfish topologies. Section~\ref{sec:traffic_profiles} defines the PQC traffic profiles and the endpoint regimes. Section~\ref{sec:routing} describes the routing strategies. Section~\ref{sec:setup} describes the experimental setup: the network instances, the cryptographic traffic classes, the experimental procedure, and the observables. Section~\ref{sec:results} presents and discusses the results, and Section~\ref{sec:conclusion} concludes.

\section{Network model and topologies}
\label{sec:topologies}

In this section we introduce the two topology families adopted in this study: the fat-tree and Jellyfish. Before describing them in detail, we also introduce the notation for network and PQC endpoint modelling. The latter follows the analytical model presented in \cite{PQCmodel}.

\subsection{Network model}
\label{sec:network_model}

A data-centre network is modelled as an undirected graph
\begin{equation}
    \mathcal{N}=(V,E,\mathbf{c}),
    \label{eq:network_def}
\end{equation}
where $V$ is the set of nodes (servers and switches), $E$ is the set of physical links, and $\mathbf{c}=(c_e)_{e\in E}$ is the vector of link capacities, with $c_e>0$ the capacity of link $e\in E$ in bits per second. A link $e=\{u,v\}$ can carry traffic in both directions, and the two directions share the same capacity $c_e$. \IP{This is an aggregate-capacity abstraction; separate directional capacities of full-duplex links are not modelled.} Every node $v\in V$ is also a potential PQC endpoint and carries the technical specification required by the endpoint model of Ref.~\cite{PQCmodel},
\begin{equation}
    \sigma_v=(N_v,f_v,\eta_v,\alpha_v,\delta_v,\rho_v),
    \label{eq:node_spec}
\end{equation}
namely the number of logical CPUs $N_v$, the clock frequency $f_v$, the CPU efficiency $\eta_v\in(0,1]$, the non-cryptographic overhead fraction $\alpha_v\geq0$, the additive per-handshake I/O delay $\delta_v\geq0$, and the parallel efficiency $\rho_v$\IP{, used here as an effective compute-capacity multiplier as explained in Table~\ref{tab:substrate}}. These parameters determine how many PQC handshakes per second a node can sustain for a given cryptographic profile (Sec.~\ref{sec:routing}). Ref.~\cite{PQCmodel} labels endpoints by $e$; here $e$ is reserved for links and nodes are labelled by $u,v$. Only the endpoints of a connection perform cryptographic work: intermediate switches forward traffic and consume link capacity but no compute. A topology family is a rule for generating $(V,E)$; a topology \textbf{instance} is one sampled graph together with sampled link capacities $\mathbf c$ and node specifications $\{\sigma_v\}_{v\in V}$. In the experiments of Sec.~\ref{sec:setup}, one instance of each family is generated once and then kept fixed.

\subsection{Fat-tree}

The fat-tree topology is a data-centre adaptation of Clos switching networks, originally introduced to construct scalable non-blocking interconnection fabrics from smaller switching elements~\cite{Clos}. Its use in modern data centres was popularised by Al-Fares, Loukissas, and Vahdat, who showed that a rearranged fat-tree built from commodity Ethernet switches can support large clusters with high aggregate bandwidth and rich path diversity~\cite{AlFares}. The central idea is to avoid the oversubscription bottleneck of a simple hierarchical tree by increasing the available capacity toward the core of the network.

\begin{figure*}[t]
\centering
\begin{tikzpicture}[
    x=0.62cm,
    y=0.68cm,
    font=\footnotesize,
    sw/.style={draw=black!70, circle, minimum size=4.8pt, inner sep=0pt, fill=white},
    core/.style={sw, fill=blue!22},
    agg/.style={sw, fill=orange!30},
    edge/.style={sw, fill=green!30},
    host/.style={draw=black!72, circle, minimum size=2.2pt, inner sep=0pt, fill=black!78},
    serveredge/.style={draw=teal!60!black, line width=0.26pt},
    edgeagg/.style={draw=black!35, line width=0.22pt},
    aggcore/.style={draw=blue!45!black, line width=0.18pt, opacity=0.45},
    podbox/.style={draw=black!35, dashed, rounded corners=1.5pt, fill=black!2, inner sep=3pt, line width=0.25pt},
    layerband/.style={draw=none, fill=black!2},
    callout/.style={draw=black!45, rounded corners=1.3pt, fill=white, inner sep=2.6pt, align=left},
    leader/.style={-{Latex[length=1.5mm]}, draw=black!50, line width=0.25pt},
    legend/.style={draw=black!35, rounded corners=1.4pt, fill=black!2, inner sep=4pt, align=left}
]
\fill[layerband] (-8.7,4.45) rectangle (8.7,5.25);
\fill[layerband] (-8.7,2.15) rectangle (8.7,2.95);
\fill[layerband] (-8.7,0.55) rectangle (8.7,1.35);
\fill[layerband] (-8.7,-0.75) rectangle (8.7,-0.05);

\foreach \i/\x in {1/-4.0,2/-3.0,3/-2.0,4/-1.0,5/0,6/1.0,7/2.0,8/3.0,9/4.0} {
    \node[core] (c\i) at (\x,4.85) {};
}

\begin{scope}[on background layer]
\foreach \p/\xbase in {1/-7.5,2/-4.5,3/-1.5,4/1.5,5/4.5,6/7.5} {
    \coordinate (pod\p-nw) at (\xbase-1.12,3.25);
    \coordinate (pod\p-se) at (\xbase+1.12,-1.05);
    \node[podbox, fit=(pod\p-nw) (pod\p-se)] {};
}
\end{scope}

\foreach \p/\xbase in {1/-7.5,2/-4.5,3/-1.5,4/1.5,5/4.5,6/7.5} {
    \foreach \j/\dx in {1/-0.72,2/0,3/0.72} {
        \node[agg] (a\p\j) at (\xbase+\dx,2.55) {};
        \node[edge] (e\p\j) at (\xbase+\dx,0.92) {};
        \foreach \m/\hx in {1/-0.22,2/0,3/0.22} {
            \node[host] (h\p\j\m) at (\xbase+\dx+\hx,-0.35) {};
            \draw[serveredge] (h\p\j\m) -- (e\p\j);
        }
    }
    \foreach \e in {1,2,3} {
        \foreach \a in {1,2,3} {
            \draw[edgeagg] (e\p\e) -- (a\p\a);
        }
    }
    \foreach \co in {1,2,3} {\draw[aggcore] (a\p1) -- (c\co);}
    \foreach \co in {4,5,6} {\draw[aggcore] (a\p2) -- (c\co);}
    \foreach \co in {7,8,9} {\draw[aggcore] (a\p3) -- (c\co);}
}

\node[callout, anchor=east] (corelabel) at (-9.35,4.85)
    {core layer\\$9$ switches};
\draw[leader] (corelabel.east) -- (c1);

\node[callout, anchor=east] (podlabel) at (-9.35,1.60)
    {six identical pods\\$3$ edge, $3$ aggregation,\\$9$ hosts per pod};
\draw[leader] (podlabel.east) -- (-8.62,3.25);

\node[callout, anchor=south] (multilabel) at (0,5.72)
    {balanced $k=6$ fat-tree\\$54$ hosts, $45$ switches, $162$ links};
\draw[leader] (multilabel.south) -- (c5);

\node[legend, anchor=west] at (8.95,1.55) {
    \begin{tabular}{@{}ll@{}}
    \tikz{\node[core] {}; } & core switch \\
    \tikz{\node[agg] {}; } & aggregation switch \\
    \tikz{\node[edge] {}; } & edge switch \\
    \tikz{\node[host] {}; } & host endpoint \\
    \tikz{\draw[serveredge] (0,0) -- (0.42,0); } & server--edge \\
    \tikz{\draw[edgeagg] (0,0) -- (0.42,0); } & edge--aggregation \\
    \tikz{\draw[aggcore] (0,0) -- (0.42,0); } & aggregation--core
    \end{tabular}
};
\end{tikzpicture}
\caption{Schematic of the $k=6$ fat-tree instance used in the fat-tree experiment. Each pod contains three edge switches, three aggregation switches, and nine hosts; the core contains $(k/2)^2=9$ switches. The $54$ hosts form the set $\mathcal A_{\mathrm{leaf}}$ used for east--west traffic; north--south traffic connects these hosts to the $9$ core switches in $\mathcal A_{\mathrm{core}}$. Aggregation--core links are drawn muted to keep the pod layout readable.}
\label{fig:fat_tree_k6}
\end{figure*}
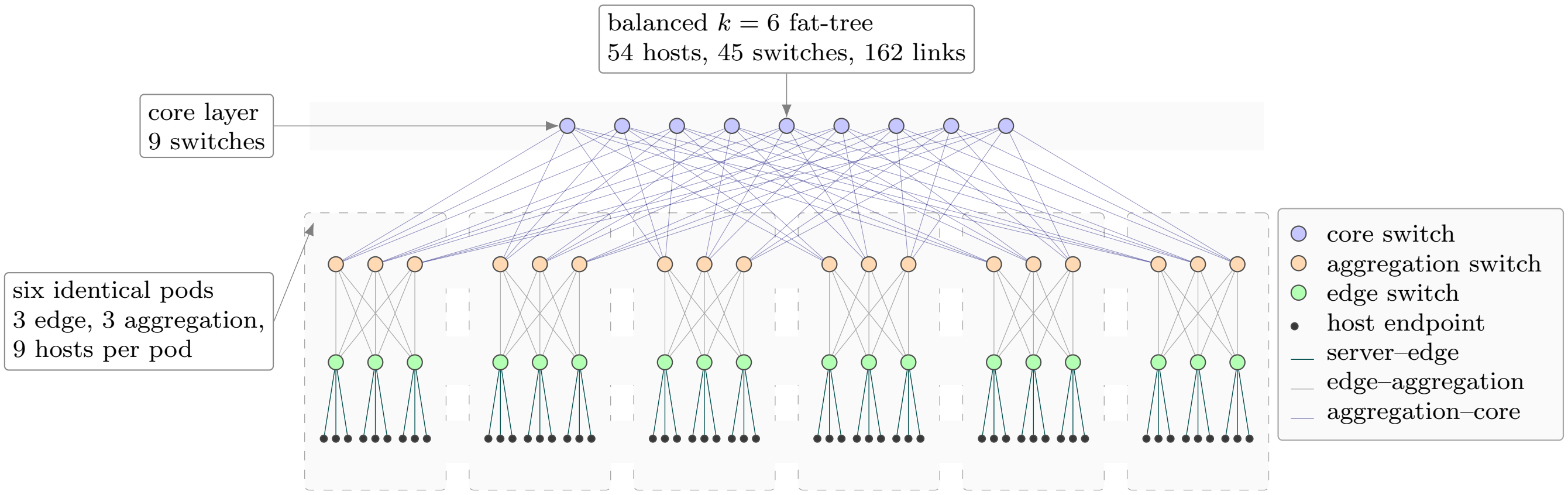

A standard $k$-ary fat-tree is built from identical $k$-port switches, where $k$ is even (the $k=6$ instance used in the experiments is shown in Fig.~\ref{fig:fat_tree_k6}). The topology is organised into $k$ pods. Each pod contains
\begin{equation}
    \frac{k}{2}~\text{edge switches}~~\text{and}~~
    \frac{k}{2}~\text{aggregation switches}.\label{eq:fatt_pod_composition}
\end{equation}
Each edge switch connects to $k/2$ hosts and to all $k/2$ aggregation switches in the same pod. The core layer contains
\begin{equation}
    \left(\frac{k}{2}\right)^2
    \label{eq:fatt_core_count}
\end{equation}
core switches. Each aggregation switch connects to $k/2$ core switches, one in each core group.
The total number of hosts is therefore
\begin{equation}
    N_{\mathrm{host}}
    =
    k \cdot \frac{k}{2} \cdot \frac{k}{2}
    =
    \frac{k^3}{4},
    \label{eq:fatt_hosts}
\end{equation}
and the total number of switches is
\begin{equation}
    N_{\mathrm{sw}}
    =
    k\left(\frac{k}{2}+\frac{k}{2}\right)
    +
    \left(\frac{k}{2}\right)^2
    =
    \frac{5k^2}{4}.
    \label{eq:fatt_switches}
\end{equation}
The number of host-facing links is $k^3/4$, the number of edge--aggregation links is $k^3/4$, and the number of aggregation--core links is also $k^3/4$. This balanced structure is one reason why fat-trees are widely used as a reference topology for scalable data-centre studies.

In a fat-tree, the relative positions of the endpoints determine the number and length of their shortest paths, where length is measured in hops. This structure helps explain the routing benefits observed in Sec.~\ref{sec:results}, because it determines the alternatives available for distributing traffic without taking longer routes. Two hosts attached to the same edge switch are joined by a unique two-hop path. Two hosts in the same pod but on different edge switches are joined by $k/2$ four-hop paths, one through each aggregation switch of the pod. Two hosts in different pods are joined by $(k/2)^2$ six-hop paths of the form edge--aggregation--core--aggregation--edge, one through each core switch. Finally, a host and a core switch are joined by a \textbf{unique shortest} three-hop path, because each core switch is reachable from a pod through exactly one aggregation switch. For $k=6$ these counts are $1$, $3$, $9$, and $1$ respectively. Longer alternatives can still be used by the unrestricted LP, and even distinct shortest paths share the hosts' access links; path diversity therefore does not remove every bottleneck. This path multiplicity makes the fat-tree suitable for evaluating policies such as OSPF-like shortest-path routing and fractional multi-path ($N$-path) routing (Sec.~\ref{sec:routing}). For the PQC benchmark, this regularity provides a controlled setting in which changes in demand coverage, congestion, and QoS can be attributed primarily to the offered traffic, routing policy, and cryptographic profile, rather than to structural irregularity.

The fat-tree therefore acts as a high-capacity reference fabric: the experiment tests how its path diversity and link capacities support PQC traffic after accounting for endpoint processing limits.

\subsection{Jellyfish}

Jellyfish was proposed by Singla, Hong, Popa, and Godfrey as an alternative data-centre topology based on random regular graphs~\cite{Jellyfish}. Instead of arranging switches into a fixed hierarchy, Jellyfish connects top-of-rack switches randomly while preserving a target inter-switch degree, i.e.\ a prescribed number of direct switch neighbours, excluding attached servers. This design was motivated by two limitations of highly structured topologies: they can be difficult to expand incrementally, and their fully populated configurations occur only at specific sizes. For example, a standard fat-tree has $k^3/4$ hosts: $54$ for $k=6$, but $128$ for the next even value, $k=8$. In contrast, Jellyfish can naturally accommodate additional switches and servers without rebuilding the entire fabric. For example, a new switch with six inter-switch ports can replace three existing links with six links to their distinct endpoints, preserving the degrees of the existing switches while adding its attached servers~\cite{Jellyfish}.

Let $n$ be the number of switches, $k$ the number of ports per switch (the same letter as the fat-tree arity, since both count the ports of a switch), and $r$ the number of ports reserved for inter-switch connectivity. Each switch then connects to
\begin{equation}
    h = k-r
    \label{eq:jelly_hosts_per_switch}
\end{equation}
hosts, giving a total number of attached servers
\begin{equation}
    N_{\mathrm{host}} = n(k-r).
    \label{eq:jelly_total_hosts}
\end{equation}
The inter-switch fabric is sampled as an approximately uniform random $r$-regular graph over $n$ switches. Hence each switch has degree $r$, and the total number of inter-switch links is
\begin{equation}
    E_{\mathrm{sw}} = \frac{nr}{2},
    \label{eq:jelly_inter_switch_links}
\end{equation}
which requires $nr$ to be even. Unlike a fat-tree, Jellyfish is not constrained by pod sizes or by the cubic host-count relation $k^3/4$. This makes it particularly attractive for incremental data-centre growth.

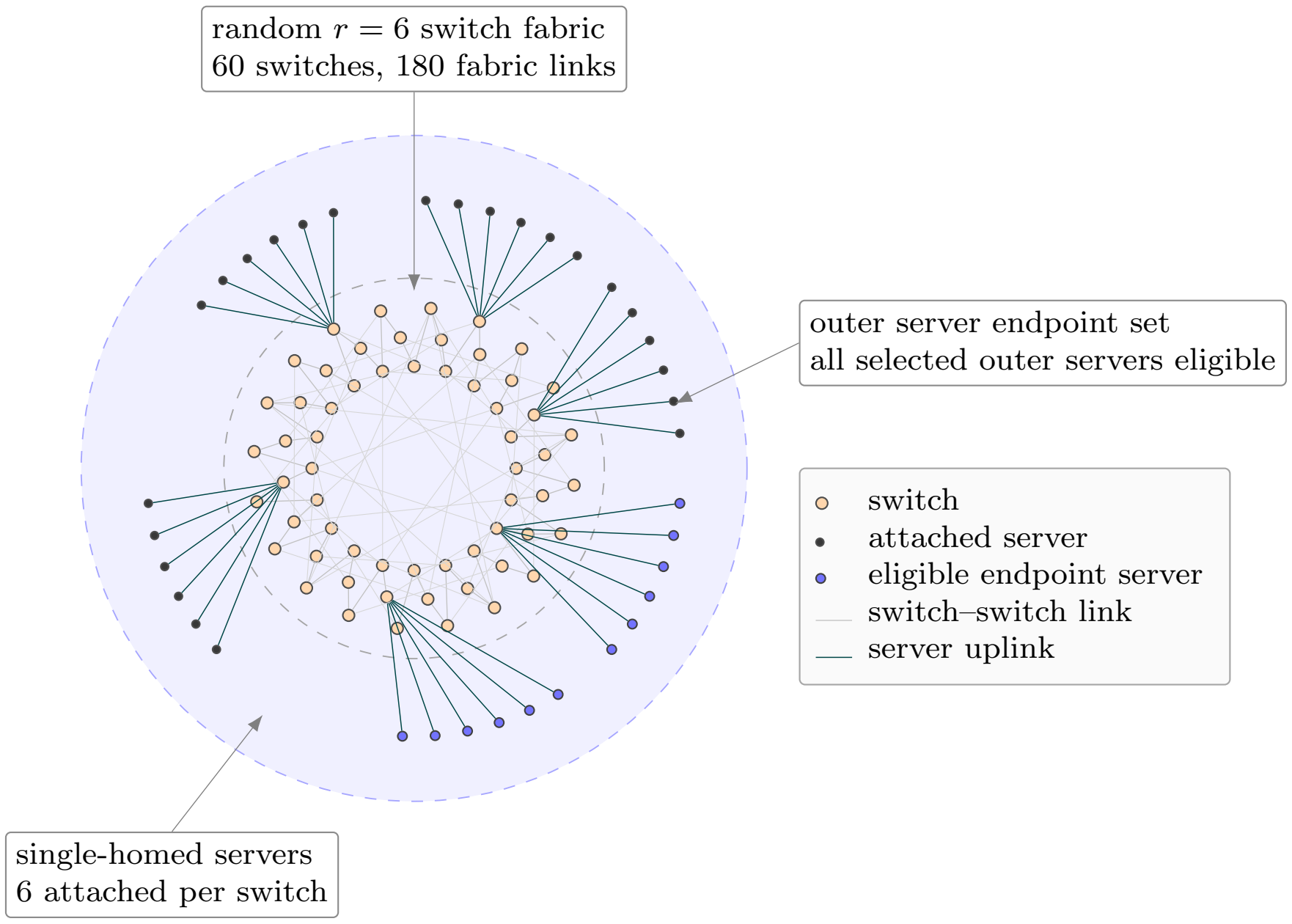
\begin{figure*}[t]
\centering
\begin{tikzpicture}[
    x=0.78cm,
    y=0.78cm,
    font=\footnotesize,
    fabric/.style={draw=black!35, dashed, line width=0.35pt},
    sw/.style={draw=black!70, circle, minimum size=3.0pt, inner sep=0pt, fill=orange!32},
    sampled/.style={draw=black!75, circle, minimum size=3.4pt, inner sep=0pt, fill=blue!30},
    host/.style={draw=black!70, circle, minimum size=2.0pt, inner sep=0pt, fill=black!78},
    endhost/.style={draw=black!75, circle, minimum size=2.4pt, inner sep=0pt, fill=blue!55},
    link/.style={draw=black!24, line width=0.25pt},
    chord/.style={draw=black!16, line width=0.20pt},
    serverlink/.style={draw=teal!55!black, line width=0.28pt},
    callout/.style={draw=black!45, rounded corners=1.3pt, fill=white, inner sep=2.6pt, align=left},
    leader/.style={-{Latex[length=1.5mm]}, draw=black!50, line width=0.25pt},
    legend/.style={draw=black!35, rounded corners=1.4pt, fill=black!2, inner sep=4pt, align=left}
]
\fill[blue!6] (0,0) circle (3.85);
\draw[fabric] (0,0) circle (2.20);
\draw[blue!35, dashed, line width=0.35pt] (0,0) circle (3.85);

\foreach \i in {1,...,60} {
    \pgfmathsetmacro{\ang}{6*\i}
    \pgfmathsetmacro{\rad}{1.18 + 0.34*mod(\i,3)}
    \pgfmathsetmacro{\x}{\rad*cos(\ang)}
    \pgfmathsetmacro{\y}{\rad*sin(\ang)}
    \node[sw] (s\i) at (\x,\y) {};
}

\foreach \i in {1,...,60} {
    \pgfmathtruncatemacro{\n}{mod(\i,60)+1}
    \draw[link] (s\i) -- (s\n);
}
\foreach \i in {1,3,...,59} {
    \pgfmathtruncatemacro{\n}{mod(\i+10,60)+1}
    \draw[chord] (s\i) -- (s\n);
}
\foreach \i in {2,6,...,58} {
    \pgfmathtruncatemacro{\n}{mod(\i+23,60)+1}
    \draw[chord] (s\i) -- (s\n);
}

\foreach \i/\baseang in {4/25,11/70,20/125,31/205,43/285,54/335} {
    \foreach \m in {1,...,6} {
        \pgfmathsetmacro{\ang}{\baseang + 7*(\m-3.5)}
        \pgfmathsetmacro{\x}{3.10*cos(\ang)}
        \pgfmathsetmacro{\y}{3.10*sin(\ang)}
        \ifnum\i>42
            \node[endhost] (h\i\m) at (\x,\y) {};
        \else
            \node[host] (h\i\m) at (\x,\y) {};
        \fi
        \draw[serverlink] (s\i) -- (h\i\m);
    }
}

\node[callout, anchor=south] (fabriclabel) at (0,4.35)
    {random $r=6$ switch fabric\\$60$ switches, $180$ fabric links};
\draw[leader] (fabriclabel.south) -- (0,2.05);

\node[callout, anchor=west] (endpointlabel) at (4.45,1.45)
    {outer server endpoint set\\all selected outer servers eligible};
\draw[leader] (endpointlabel.west) -- (3.02,0.75);

\node[callout, anchor=north] (serverlabel) at (-2.8,-4.20)
    {single-homed servers\\$6$ attached per switch};
\draw[leader] (serverlabel.north) -- (-1.75,-2.85);

\node[legend, anchor=west] at (4.45,-1.25) {
    \begin{tabular}{@{}ll@{}}
    \tikz{\node[sw] {}; } & switch \\
    \tikz{\node[host] {}; } & attached server \\
    \tikz{\node[endhost] {}; } & eligible endpoint server \\
    \tikz{\draw[link] (0,0) -- (0.42,0); } & switch--switch link \\
    \tikz{\draw[serverlink] (0,0) -- (0.42,0); } & server uplink
    \end{tabular}
};
\end{tikzpicture}
\caption{Illustrative schematic of the Jellyfish topology used in the experiment. The instance has $60$ switches with $12$ ports and $6$ attached servers per switch, leaving $r=6$ inter-switch ports and $180$ switch--switch links. Only representative fabric links and server attachments are drawn. The outer server endpoint set $\mathcal A_{\mathrm{out}}$, selected by the outer-node rule, contains $85$ servers attached to $16$ switches in the experimental instance; every server in this set is eligible for end-to-end demand sampling.}
\label{fig:jellyfish_60}
\end{figure*}

For fixed $r>2$, uniformly sampled random $r$-regular graphs have a diameter, and hence an average shortest-path length, of order $\log n$ with high probability~\cite{BollobasVega}. Here, ``with high probability'' means that the probability over graph realisations tends to one as $n\to\infty$; it is an asymptotic statement, not a guarantee for every finite instance. The typical switch-to-switch distance scales approximately as
\begin{equation}
    {\bar h} \sim \frac{\log n}{\log(r-1)}.
    \label{eq:jelly_path_length}
\end{equation}
The instance used in this paper (Fig.~\ref{fig:jellyfish_60}) has $n=60$ switches with $k=12$ ports, of which $r=6$ connect to other switches and $h=6$ to servers, giving $360$ servers and $180$ switch--switch links. Its measured mean switch-to-switch distance is $\bar h=2.45$ hops, with a maximum of $4$, close to the estimate $\log 60/\log 5\approx2.54$ of Eq.~(\ref{eq:jelly_path_length}). A server-to-server path therefore has on average about $4.5$ hops, between the four-hop intra-pod and six-hop inter-pod paths of the fat-tree.

Thus, even large Jellyfish networks can provide short paths and substantial path diversity. However, this diversity is irregular: not all source--destination pairs see the same number or quality of candidate paths. Routing therefore plays a more central role than in a symmetric fat-tree. The achieved performance depends on how effectively the routing algorithm exploits the available random graph structure.

For the PQC benchmark, Jellyfish therefore provides the irregular counterpart to the fat-tree. The same traffic-generation and compute workflow can be evaluated on a fabric where path diversity is present but unevenly distributed across endpoint pairs. This makes it possible to test whether changes in demand coverage, congestion, and QoS are driven by the cryptographic profile itself, or by the routing policy's ability to exploit a random and locally heterogeneous topology.

\section{Traffic profiles}
\label{sec:traffic_profiles}

The traffic profile defines the end-to-end communication demands evaluated on a topology. Each demand represents a PQC-protected connection between two selected endpoints. The topology determines which endpoint pairs and paths are available; the traffic profile determines which pairs are active, how much load they request, and which cryptographic configuration is attached to each connection.
We model a traffic profile as a set
\begin{equation}
    \mathcal{D} = \{d_i\}_{i=1}^{M},
    \label{eq:traffic_profile_set}
\end{equation}
where $M$ is the number of generated end-to-end connections. We call one sampled set $\mathcal D$ a \textbf{traffic matrix}; the rule that generates it, i.e.\ the endpoint regime and the laws of the rates and profiles introduced below, is the \textbf{traffic profile}. Each demand is specified by
\begin{equation}
    d_i = (s_i,t_i,\lambda_i,\kappa_i),
    \label{eq:demand_record}
\end{equation}
where $s_i$ and $t_i$ are the source and destination endpoints, $\lambda_i$ is the requested rate, and $\kappa_i$ is the PQC configuration assigned to the connection. \IP{Each demand record represents a stream of new handshakes between one endpoint pair, and $M$ counts these demand streams.} The source plays the TLS client role and the destination the TLS server role; this matters because the two roles execute different cryptographic operations (Sec.~\ref{sec:routing}). Requested rates are measured in connections per second (conn/s), i.e.\ in completed handshakes per second; we also use kconn/s, with $1~\mathrm{kconn/s}=10^3$~conn/s. Throughout the paper, $\mathcal{U}_{\mathbb{Z}}(a,b)$ denotes the discrete uniform distribution over the integers $\{a,a+1,\ldots,b\}$, used for quantities that are integer-valued by construction (e.g.\ connection counts, logical-core counts); $\mathcal{U}(a,b)$ denotes the continuous uniform distribution on $[a,b]$, used for real-valued quantities. In the implemented use cases, $\lambda_i$ is sampled from a configurable distribution, typically
\begin{equation}
    \lambda_i \sim \mathcal{U}_{\mathbb{Z}}
    \left(\lambda_{\min},\lambda_{\max}\right)~{\mathrm{conn/s}},
    \label{eq:lambda_dist_general}
\end{equation}
i.e.\ the requested rate is drawn as an integer number of connections per second.
The PQC configuration is represented by
\begin{equation}
    \kappa_i =
    \left(
        \mathrm{SIG}_i,
        \mathrm{KEM}_i,
        \mathrm{SYM}_i,
        \ell_i
    \right),
    \label{eq:kappa_def}
\end{equation}
where $\mathrm{SIG}_i$ is the digital signature algorithm, $\mathrm{KEM}_i$ is the key-encapsulation mechanism, $\mathrm{SYM}_i$ is the symmetric authenticated-encryption primitive, and $\ell_i$ is the assigned NIST security level. The detailed endpoint and network performance model used to associate node capabilities, cryptographic costs, and link-level quantities is described in Ref.~\cite{PQCmodel}. Here we only specify the traffic-generation parameters used to instantiate the use cases.

The cryptographic profiles considered in the experiments are aligned with the NIST post-quantum cryptography standards: ML-KEM is standardised in FIPS~203~\cite{FIPS203}, ML-DSA in FIPS~204~\cite{FIPS204}, and SLH-DSA in FIPS~205~\cite{FIPS205}. These standards provide the reference families for post-quantum key establishment and authentication, while the benchmark framework allows different parameter sets and security levels to be assigned to different generated connections. In addition to ML-KEM, the experiments use Classic McEliece~\cite{ClassicMcEliece} as the key-encapsulation mechanism of a deliberately heavy traffic class. Classic McEliece is a conservative code-based scheme with very small ciphertexts but public keys of $0.5$--$1.4$~MB. Such keys exceed the $2^{16}-1$ byte limit of the \texttt{key\_share} extension of TLS~1.3~\cite{RFC8446}, so Classic McEliece cannot be negotiated as an ephemeral key share in standard TLS~1.3 and is normally proposed for long-lived static keys or for KEM-based authentication~\cite{KEMTLS}. We nevertheless model it as if it were carried in the handshake, because it provides a \IP{large-key stress scenario} for the network-level cost of large-key PQC: the byte budget of a Classic McEliece handshake is \IP{about seventy times larger on average} than that of an ML-KEM handshake (Sec.~\ref{sec:traffic_classes}), while its signature and symmetric primitives are unchanged.

Each traffic profile is generated for an \textbf{endpoint regime}, i.e.\ a rule that fixes the admissible endpoint set $\mathcal A\subseteq V$ from which the pairs $(s_i,t_i)$ are drawn. The three regimes used in this paper are described next; the numerical sizes of the corresponding endpoint sets are given in Sec.~\ref{sec:setup}.

\subsection{East--west traffic}

East--west traffic models communication between endpoints at the same logical level of the data-centre fabric. In cloud systems this corresponds to internal service-to-service communication, including microservices, virtual machines, containers, storage nodes, distributed databases, and replicated application components, and can account for a large share of network traffic. Measurements by Benson et al.~\cite{Benson2010} show that traffic locality varies across deployments, with most traffic remaining within a rack in the cloud data centres they studied.

In the fat-tree use case, east--west traffic is generated by selecting both endpoints from the leaf or server-facing layer{,
\begin{equation}
    s_i,t_i\in\mathcal{A}_{\mathrm{leaf}},
    \qquad s_i\neq t_i,
    \label{eq:eastwest_endpoints}
\end{equation}
where $\mathcal A_{\mathrm{leaf}}$ is the set of all hosts.} This profile stresses the internal fabric because flows may remain local or traverse aggregation and core links depending on endpoint placement.

\subsection{North--south traffic}

North--south traffic models communication between lower-level endpoints and higher-level infrastructure nodes, such as core, gateway, or external-facing services. In cloud deployments this represents client-to-service traffic, ingress and egress flows, API access, and communication with shared control-plane or gateway resources.

In the fat-tree use case, north--south traffic is generated by selecting one endpoint from the leaf/server-facing layer and the other from the root or core layer{,
\begin{equation}
    \{s_i,t_i\}=\{u,v\}
    \quad\text{with}\quad
    u\in\mathcal{A}_{\mathrm{leaf}},\;
    v\in\mathcal{A}_{\mathrm{core}},
    \label{eq:northsouth_endpoints}
\end{equation}
where $\mathcal A_{\mathrm{core}}$ is the set of core switches, which act as PQC endpoints in this regime.} The direction can be sampled randomly or fixed as leaf-to-root or root-to-leaf; in the experiments it is sampled uniformly at random, so that the client role is assigned to the host or to the core node with equal probability. This profile is relevant for studying PQC migration on access and gateway paths. A host and a core switch have a unique shortest three-hop path (Sec.~\ref{sec:topologies}), although longer alternatives exist. The consequences of this restricted shortest-path diversity are examined in Sec.~\ref{sec:results}.

\subsection{Jellyfish traffic}

Jellyfish does not provide a natural leaf--aggregation--core hierarchy. For this reason, the traffic profile uses a single outer-server endpoint regime rather than an east--west or north--south classification. {The generator embeds the random graph in the plane and ranks all nodes by their distance from the centre of the layout; \IP{the \textbf{outer-node rule} uses a nominal quantile parameter of $0.8$ to select approximately the outermost $20\%$ of nodes as server endpoints.} This defines the admissible endpoint set $\mathcal A_{\mathrm{out}}$, and every Jellyfish demand satisfies
\begin{equation}
    s_i,t_i\in\mathcal{A}_{\mathrm{out}},
    \qquad s_i\neq t_i.
    \label{eq:jelly_endpoints}
\end{equation}
All nodes in this set are eligible endpoints, rather than a further sampled subset. The rule serves two purposes. It reproduces the locality of a service cluster confined to a group of racks, and it concentrates the load on the fabric ports of those racks instead of spreading it uniformly over all switches, which would distribute the offered traffic across more fabric ports at the same $M$. The \IP{nominal} quantile $0.8$ is a compromise between this concentration and having enough endpoints for the largest traffic matrices. Only the selected outer servers perform cryptographic work in the endpoint model; their processing parameters are specified in Sec.~\ref{sec:instances}.}

This provides a consistent way to generate end-to-end server--server demands on an irregular random graph while encouraging traffic to traverse the internal Jellyfish fabric: the outer servers are attached to a minority of the switches, so most sampled pairs are served by different top-of-rack switches and must cross the random fabric. The detailed numerical configuration used in the two experiments is given in Sec.~\ref{sec:setup}.

\section{Routing strategies}
\label{sec:routing}

Routing determines how the traffic matrices introduced above are embedded into the selected data-centre topology. We denote a routing policy by $\pi$; three policies are considered. The policies considered here differ in two main respects: whether routes are selected independently or with knowledge of the complete traffic matrix, and whether the allocation of each connection is retained on one path or distributed across a bounded number of paths.

{
\subsection{From handshake rates to bandwidth demands}
\label{sec:demand_conversion}

Consider the network $\mathcal{N}=(V,E,\mathbf{c})$ of Sec.~\ref{sec:network_model} and a traffic matrix $\mathcal D$ of Sec.~\ref{sec:traffic_profiles}. The connection record $d_i=(s_i,t_i,\lambda_i,\kappa_i)$ carries the requested rate $\lambda_i$ in conn/s, whereas link capacities are expressed in bit/s. Before comparing routing policies, we first determine the handshake rates that the endpoints can sustain and convert these into bandwidth demands. After routing, the allocated bandwidth is converted back into satisfied handshake rates. The following three steps summarise this procedure, based on the model of Ref.~\cite{PQCmodel}.

\textbf{Step 1: Endpoint feasibility.} Each handshake of connection $i$ requires a fixed number of CPU cycles at the client $s_i$ (KEM key generation, KEM decapsulation, signature verification) and at the server $t_i$ (KEM encapsulation, signature generation), determined by the profile $\kappa_i$. \IP{Following Ref.~\cite{PQCmodel}, this workload assumes server authentication with a public key already trusted by the client; certificate-chain validation and client authentication are excluded.} Given the specification $\sigma_v$ of Eq.~(\ref{eq:node_spec}), an endpoint $v$ can sustain a bounded aggregate handshake rate. When the connections incident on $v$ request more compute than $v$ can provide, their rates are proportionally rescaled in the compute domain, and the endpoint-feasible rate of connection $i$ is the minimum over its two endpoints,
\begin{equation}
    \lambda_i^{\mathrm{ep}}
    =
    \min\{\lambda^{\prime}_{s_i,i},\lambda^{\prime}_{t_i,i}\}
    \leq\lambda_i .
    \label{eq:endpoint_feasible_rate}
\end{equation}
This step depends only on $\mathcal D$ and on the node specifications, not on the routing policy. The original request $\lambda_i$ is retained as the reference for measuring service; the adjusted rate $\lambda_i^{\mathrm{ep}}$ is used to construct the bandwidth demand passed to routing.

\textbf{Step 2: Demand conversion.} A handshake under profile $\kappa_i$ transmits a fixed byte budget $B_i$, the sum of a protocol-dependent constant and of the sizes of the public key, ciphertext, certificate, and signature of the selected primitives. The routing demand is
\begin{equation}
    b_i=\frac{8\,B_i\,\lambda_i^{\mathrm{ep}}}{1-p},
    \label{eq:demand_conversion}
\end{equation}
where $p$ is the packet-loss probability ($p=0$ in this paper). Table~\ref{tab:byte_budget} lists $B_i$ for the profiles used in the experiments.

\textbf{Step 3: Routing and back-conversion.} The routing policy assigns to connection $i$ an allocated bandwidth $a_i$ with $0\le a_i\le b_i$ on one or more $s_i$--$t_i$ paths. The satisfied rate and the service ratio are then
\begin{equation}
    r_i=\frac{a_i\,(1-p)}{8\,B_i},
    \qquad
    q_i=\frac{r_i}{\lambda_i}
    =
    \frac{a_i}{b_i}\cdot\frac{\lambda_i^{\mathrm{ep}}}{\lambda_i}.
    \label{eq:back_conversion}
\end{equation}
The allocation $a_i$ is carried on one or more paths, and we denote by $f_{i,e}\ge0$ the resulting flow of connection $i$ on link $e$, summed over its paths, so that $f_{i,e}=a_i$ on every link of a single-path route. The factorisation of $q_i$ separates an endpoint factor, common to all routing policies, from a network factor that the policy controls. In the rest of this section we describe how $a_i$ and the flows $f_{i,e}$ are obtained.
}

\subsection{Shortest-path and OSPF-like routing}

A \textbf{single-unicast} routing problem specifies one source $s_i$, one destination $t_i$, and one end-to-end demand $b_i$. Shortest-path routing assigns this connection to a path of minimum additive cost. If $\mathcal{P}_i$ is the set of paths from $s_i$ to $t_i$ and $w_e$ is the cost of link $e$, the selected path is
\begin{equation}
    P_i^\star
    =
    \arg\min_{P\in\mathcal{P}_i}
    \sum_{e\in P}w_e.
    \label{eq:shortest_path_general}
\end{equation}
This family is rooted in Dijkstra's algorithm~\cite{Dijkstra} and forms the basis of link-state routing protocols such as OSPF~\cite{OSPF}. In the OSPF-like policy used in the present experiments, the link cost is the inverse of the fixed link capacity,
\begin{equation}
    w_e=\frac{1}{c_e},
    \qquad
    P_i^\star
    =
    \arg\min_{P\in\mathcal{P}_i}
    \sum_{e\in P}\frac{1}{c_e}.
    \label{eq:ospf_path}
\end{equation}
Thus high-capacity links have lower cost, but the objective remains an additive least-cost objective rather than a widest-path objective.

For a fixed source $s_i$, Dijkstra's algorithm assigns each node $u\in V$ a tentative label $T_u$, with $T_{s_i}=0$ and $T_u=\infty$ for $u\neq s_i$. When a settled node $u$ is used to inspect a neighbour $v$, the label is relaxed according to
\begin{equation}
    T_v
    \gets
    \min\left\{
        T_v,\,
        T_u+\frac{1}{c_{uv}}
    \right\}.
    \label{eq:ospf_relaxation}
\end{equation}
If the second term is smaller, $u$ becomes the predecessor of $v$. With a binary heap, the search runs in $\mathcal{O}((|E|+|V|)\log|V|)$ time. Once $P_i^\star$ has been reconstructed, the independently admissible allocation is
\begin{equation}
    \widetilde a_i
    =
    \min\left\{
        b_i,\,
        \min_{e\in P_i^\star}c_e
    \right\}.
    \label{eq:ospf_independent_rate}
\end{equation}
The implemented policy retains one predecessor path and does not split equal-cost traffic across several paths. We therefore use the term OSPF-like: the policy captures capacity-weighted link-state path selection, but not the full OSPF control plane or equal-cost multi-path forwarding~\cite{ECMP}. It is intended as a deliberately simple baseline that uses no information about the traffic matrix, against which the value of that information can be measured. \IP{In a fat-tree, several paths between two hosts can have the same hop count, while their inverse-capacity costs can differ because link capacities are sampled. Independently selected least-cost paths may share aggregation and core links. When weighted costs tie, the implementation retains the path selected by its deterministic traversal order.}

The construction so far treats one endpoint pair in isolation and therefore describes single-unicast routing. The benchmark instead considers $M$ simultaneous connections. We call this a \textbf{multiple-unicast} problem and index the connections by $\mathcal I:=\{1,\ldots,M\}$, where connection $i$ has source $s_i$, destination $t_i$, and demand $b_i$. The connections remain logically independent but compete for the same physical-link capacities. This differs from multicast, in which several destinations receive a common information flow from one source.

To obtain a multiple-unicast allocation from the OSPF-like policy, we first apply Eqs.~(\ref{eq:ospf_path}) and~(\ref{eq:ospf_independent_rate}) independently to every connection. This gives the path-rate pair $(P_i^\star,\widetilde a_i)$ for connection $i$. Although each pair is feasible in isolation, their simultaneous use can overload a link shared by several paths. We therefore use an \textbf{overlay-slice} method to reconcile the allocations while keeping every selected path unchanged. The procedure has three stages.

\textbf{(i) Identify shared links and collisions.} Initialise $a_i\gets\widetilde a_i$. For each link $e$, define the connections whose paths contain $e$, their current aggregate load, and the set of shared links as
\begin{equation}
\begin{aligned}
    I_e&:=\{i\in\mathcal I:e\in P_i^\star\},
    &L_e&:=\sum_{i\in I_e}a_i,\\
    \mathcal X&:=\{e\in E:|I_e|>1\}.
\end{aligned}
    \label{eq:overlay_collision}
\end{equation}
A shared link $e\in\mathcal X$ is a collision precisely when $L_e>c_e$. Equality is feasible. Since every path-rate pair is individually feasible, a capacity violation can occur only on a link in $\mathcal X$. The complete allocation is concurrent if and only if
\begin{equation}
    L_e\leq c_e,
    \qquad \forall e\in E.
    \label{eq:concurrency}
\end{equation}

\textbf{(ii) Iterate over the shared-link set.} For each $e\in\mathcal X$, recompute $L_e$ using the current rates. If $L_e>c_e$, proportionally rescale every connection in $I_e$ according to
\begin{equation}
    a_i\gets\frac{c_e}{L_e}a_i,
    \qquad \forall i\in I_e.
    \label{eq:overlay_slice}
\end{equation}
Because connection $i$ uses one path at the constant rate $a_i$, changing $a_i$ updates its allocation on every link of $P_i^\star$ and preserves flow conservation. The common factor preserves the relative rates of the connections involved in the collision. Each update only decreases rates. Consequently, a processed link remains feasible and no new collision can be created, so one pass over $\mathcal X$ is sufficient. The final allocation may depend on the order in which the links of $\mathcal X$ are visited; the implementation uses a fixed order, so that the result is reproducible.

\textbf{(iii) Validate concurrency.} After all links in $\mathcal X$ have been processed, the implementation checks Eq.~(\ref{eq:concurrency}) for every shared link, up to numerical tolerance. The allocation is accepted only if this validation succeeds. Overlay slicing is a local proportional reconciliation rather than a joint traffic-engineering optimisation. It restores concurrency but does not revise paths or reallocate released capacity elsewhere in the network.

\subsection{Fractional multi-commodity-flow routing}

A \textbf{multi-commodity-flow (MCF)} formulation represents every connection in the multiple-unicast index set $\mathcal I$ as a commodity~\cite{Ahuja,PioroMedhi}. Commodity $i$ has source $s_i$, destination $t_i$, and routing demand $b_i$. Let $x_{uv}^i\geq0$ be its directed flow from $u$ to $v$, and let $a_i$ be its admitted allocation. Because the two directions of a physical link $e=\{u,v\}$ share the same capacity, the link constraint is
\begin{equation}
    \sum_{i\in\mathcal I}
    \left(x_{uv}^i+x_{vu}^i\right)
    \leq c_e,
    \qquad e=\{u,v\}\in E.
    \label{eq:mcf_capacity}
\end{equation}
For each commodity, flow conservation requires
\begin{equation}
    \sum_{v:\{u,v\}\in E}
    \left(x_{uv}^i-x_{vu}^i\right)
    =
    \begin{cases}
        a_i, & u=s_i,\\
        -a_i, & u=t_i,\\
        0, & \text{otherwise},
    \end{cases}
    \label{eq:mcf_conservation}
\end{equation}
with
\begin{equation}
    0\leq a_i\leq b_i.
    \label{eq:mcf_demand_bound}
\end{equation}
A feasible solution of Eqs.~(\ref{eq:mcf_capacity})--(\ref{eq:mcf_demand_bound}) defines \textbf{fractional routing}: the variables are continuous, so the allocation of one commodity may be divided among several paths while all commodities are optimised concurrently. An \textbf{unrestricted fractional} solution imposes no upper bound on the number of paths carrying positive flow. Since the variables, constraints, and objectives are continuous and linear, the problem is a linear programme (LP). The two LP stages described below are solved with HiGHS, an open-source high-performance solver for large sparse linear programmes~\cite{HiGHS}.

The implemented fractional-routing algorithm uses two LP stages. The \textbf{first stage} maximises the minimum routing service fraction,
\begin{equation}
    \max_{\mathbf{x},\mathbf{a},\widehat q}\ \widehat q
    \qquad
    \text{subject to}
    \qquad
    a_i\geq \widehat q\,b_i,
    \quad \forall i\in\mathcal I,
    \label{eq:mcf_maxmin}
\end{equation}
together with Eqs.~(\ref{eq:mcf_capacity})--(\ref{eq:mcf_demand_bound}). This prevents the optimisation from increasing aggregate allocation by starving a difficult commodity. If the first stage returns $a_i^{(1)}$, define the achieved fraction of commodity $i$ as $\widehat q_i^{(1)}=a_i^{(1)}/b_i$. The \textbf{second stage} then solves
\begin{equation}
    \max_{\mathbf{x},\mathbf{a}}\ \sum_{i\in\mathcal I}a_i
    \qquad
    \text{subject to}
    \qquad
    a_i\geq \widehat q_i^{(1)}b_i,
    \quad \forall i\in\mathcal I,
    \label{eq:mcf_maxthroughput}
\end{equation}
under the same capacity, conservation, and demand constraints. The first stage establishes the service floor. The second uses any remaining capacity to increase the total \IP{allocated bandwidth} without reducing, up to numerical tolerance, the service achieved by any commodity in the first solution.

The LP returns an edge-flow description. To obtain explicit routes, the implementation applies a greedy residual-flow decomposition: it repeatedly extracts a positive-flow path from $s_i$ to $t_i$, assigns the bottleneck residual allocation to that path, and subtracts it from the residual flow. \IP{A feasible LP flow admits a decomposition into positive source--destination path flows and a residual circulation. Denoting the path set by $\mathcal P_i^+$, the decomposition satisfies, up to numerical tolerance,}
\begin{equation}
    x_{uv}^{i,\star}
    =
    \sum_{\substack{p\in\mathcal{P}_i^+\\(u,v)\in p}}
    f_{i,p}\IP{{}+z_{uv}^{i}},
    \qquad
    a_i^\star
    =
    \sum_{p\in\mathcal{P}_i^+}f_{i,p},
    \label{eq:mcf_path_decomposition}
\end{equation}
where $f_{i,p}>0$ is the allocation carried by path $p$. \IP{Here $z^i\geq0$ is a circulation, with zero net flow at every node, and therefore contributes no admitted source--destination rate.}
Here, $x_{uv}^{i,\star}$ is the total flow across the edge $(u,v)$ associated with commodity $i$, while $a_i^\star$ is the total allocation to $i$.
\IP{The complete set of path components realises the admitted allocations of the unrestricted fractional solution.} To impose a path-count bound, the extracted components are ordered by allocation,
\begin{equation}
    f_{i,p_1}\geq f_{i,p_2}\geq\cdots\geq f_{i,p_{{n_i}}},
\end{equation}
where ${n_i}=|\mathcal P_i^+|$. \textbf{Fractional $N$-path routing} retains the largest $N$ components. With $\chi_{uv}^{p}=\mathbf{1}\{(u,v)\in p\}$, the retained allocation is
\begin{equation}
\begin{aligned}
    x_{uv}^{i,(N)}
    &={}
    \sum_{j=1}^{\min\{N,{n_i}\}}
    f_{i,p_j}\chi_{uv}^{p_j},\\
    a_i^{(N)}
    &={}
    \sum_{j=1}^{\min\{N,{n_i}\}}f_{i,p_j}.
\end{aligned}
    \label{eq:mcf_k_path}
\end{equation}
Retaining complete path components preserves flow conservation, and removing the other components preserves the capacity constraints of the unrestricted solution. The discarded allocation is not reassigned by a subsequent optimisation. The $N$-path restriction is therefore a greedy post-processing heuristic and need not produce the optimal allocation among all routes using at most $N$ paths. \IP{The minimum-service guarantee applies to the unrestricted solution and the endpoint-adjusted demands; retaining a bounded number of paths can reduce the resulting service fractions.}

Both optimisation-based policies in the benchmark begin from the unrestricted fractional LP solution and instantiate Eq.~(\ref{eq:mcf_k_path}) at a fixed $N$. \textbf{Fractional $1$-path routing} sets $N=1$ and retains the dominant component of each decomposed flow. \textbf{Fractional $N$-path routing} denotes the general case $N>1$; the policy used in this paper is \textbf{fractional $5$-path routing}, with $N=5$. In these names, ``fractional'' identifies how the original allocation is computed, while ``$1$-path'' or ``$N$-path'' identifies the retained output; we avoid the informal labels ``single-path'' and ``multi-path'' for these two policies except as a plain-language reminder; e.g., fractional $1$-path routing is a fractional single-path policy, and fractional $5$-path routing is a fractional multi-path policy.

Comparing OSPF-like routing with the fractional policies examines the benefit of globally coordinated allocation, while comparing fractional $1$-path and $5$-path routing shows the additional demand served by retaining more paths. These benefits depend on endpoint placement: host--core pairs have a unique shortest path, although longer alternatives remain available to the LP. Section~\ref{sec:results} examines these effects across the endpoint regimes.

\section{Experimental setup}
\label{sec:setup}

The experiments examine how much post-quantum handshake demand the two networks can serve as the offered load increases, and how this depends on the cryptographic traffic class and routing policy. We generate one fat-tree and one Jellyfish instance, each represented by the network $\mathcal N=(V,E,\mathbf c)$ and node specifications $\{\sigma_v\}_{v\in V}$ introduced in Sec.~\ref{sec:network_model}. For each instance, node parameters and link capacities are drawn from the distributions specified below. The prescribed ranges allow different values across nodes and links; once sampled, these values and the connectivity are retained for every evaluation on that instance. Routing policies therefore face the same network resources within each experiment.

On each network, we sample traffic matrices $\mathcal D=\{d_i\}_{i=1}^M$ using the endpoint rules and cryptographic profiles described above. Each record $d_i=(s_i,t_i,\lambda_i,\kappa_i)$ specifies which endpoints communicate, their requested handshake rate, and the cryptographic configuration they use. Following Sec.~\ref{sec:demand_conversion}, we account for endpoint processing limits and convert the requested rates into bandwidth demands. Each routing policy $\pi$ then assigns bandwidth $a_i$ to the connections and determines the flows $f_{i,e}$ on the links. These allocations tell us how much of the requested demand is served and how much network capacity it consumes.

We repeat this process for both traffic classes and each endpoint regime, increasing the number of demand records $M$. Within each class, the rate ranges and profile probabilities stay the same, while new endpoint pairs, rates, and profiles are sampled for every traffic matrix. Thus, network parameters are sampled once per instance, whereas traffic is sampled anew in each repeat. Each sampled matrix is then evaluated under all three routing policies. The following subsections give the network parameters, traffic classes, sampling procedure, and measures used to compare the outcomes.

\subsection{Network instances and substrate parameters}
\label{sec:instances}

We first describe the size and connectivity of the two networks, then specify the endpoint processing resources and link capacities. Table~\ref{tab:substrate} gives the distributions used to sample these resources.

{
\textbf{Fat-tree instance.} The fat-tree experiment uses a $k=6$ fat-tree (Fig.~\ref{fig:fat_tree_k6}), i.e.\ a graph with $|V|=99$ nodes ($54$ hosts, $18$ edge switches, $18$ aggregation switches, $9$ core switches) and $|E|=162$ links, $54$ in each of the three layers. The endpoint sets of Eqs.~(\ref{eq:eastwest_endpoints}) and~(\ref{eq:northsouth_endpoints}) are $\mathcal A_{\mathrm{leaf}}\subset V$, the $54$ hosts, and $\mathcal A_{\mathrm{core}}\subset V$, the $9$ core switches. Every host has a single link to its edge switch, and every core switch has $6$ links, one to each pod.

\textbf{Jellyfish instance.} The Jellyfish experiment uses a random $r=6$-regular fabric of $n=60$ switches with $k=12$ ports and $h=6$ servers each (Fig.~\ref{fig:jellyfish_60}): $|V|=420$ nodes ($360$ single-homed servers and $60$ switches) and $|E|=540$ links ($180$ switch--switch links and $360$ server uplinks). The outer-node rule with \IP{nominal} quantile $0.8$ selects $\mathcal A_{\mathrm{out}}\subset V$, a set of $85$ servers \IP{(about $20.2\%$ of all $420$ nodes)} attached to $16$ of the $60$ switches, so that on average about $5$ eligible endpoints share one top-of-rack switch. All remaining nodes, including every switch, form the inner region.

\textbf{Endpoint specifications.} The parameters $\sigma_v$ of Eq.~(\ref{eq:node_spec}) determine the handshake rate that an endpoint can sustain under the model of Ref.~\cite{PQCmodel}. In the fat-tree, the eligible endpoints are hosts and, for north--south traffic, core switches. In Jellyfish, only the selected outer servers in $\mathcal A_{\mathrm{out}}$ act as endpoints and perform cryptographic work; the switches forward traffic without incurring a cryptographic compute cost. The endpoint parameters are sampled once from the distributions in Table~\ref{tab:substrate} and retained across all traffic evaluations. The same distributions apply to eligible endpoints in both topologies, with $f_v\sim\mathcal U(3.5,5.5)$~GHz. The per-primitive CPU-cycle counts were measured on the AMD EPYC 7763 reference host and are scaled by $1/(f_v\eta_v)$ in the endpoint model. Sampling clock frequencies and logical-core counts allows the simulated endpoints to have different processing capacities. \IP{The sampled specifications provide model estimates of effective cryptographic capacity, with high CPU-efficiency factors representing favourable processing conditions; they are not measured hardware configurations.}

Connections sharing an endpoint also share its processing capacity. Before routing, the model reduces their requested handshake rates where necessary to respect endpoint limits (Sec.~\ref{sec:demand_conversion}). All three routing policies receive the same adjusted demands, ensuring a consistent comparison.

\textbf{Link capacities.} The capacity $c_e$ of a link is drawn from the distribution of its group: for the fat-tree, the layer of the link; for Jellyfish, server uplinks are identified structurally as links incident to a degree-one server node and the remaining links form the random switch--switch fabric. In the fat-tree, each edge switch has three links to hosts and three to aggregation switches. The combined host-link capacity is about $1.6$ times the combined capacity of the aggregation links on average, corresponding to an access-layer \textbf{oversubscription ratio} of about $1.6{:}1$. This choice reflects the use of oversubscription to reduce network provisioning costs: Al-Fares et al.~\cite{AlFares} describe hierarchical data-centre designs with overall oversubscription ratios of $2.5{:}1$ to $8{:}1$. Such oversubscription can make traffic leaving the edge switch encounter a bottleneck even when the host links have spare capacity. For Jellyfish, each server uplink represents several physical interfaces combined into one logical link. Its larger capacity reduces the likelihood of a bottleneck at the server attachment, allowing the experiment to examine how connections compete for capacity inside the switch fabric.

The two deployments differ in both size and available capacity: Jellyfish has more switches and servers, and its server uplinks have about six times the capacity of the fat-tree host links on average. The results in Sec.~\ref{sec:results} should therefore be read as the performance of these particular deployments. Differences between them reflect the available resources and endpoint placement as well as the topology; higher demand coverage alone does not establish that one topology is intrinsically better.
}

\begin{table*}[t]
\caption{Parameter distributions used to generate the fixed network instances. Left: endpoint specification $\sigma_v$ of Eq.~(\ref{eq:node_spec}); the listed distributions apply to eligible PQC endpoints in both topologies. For Jellyfish, these are the outer servers in $\mathcal A_{\mathrm{out}}$. All sampled values are held fixed across evaluations. Ref.~\cite{PQCmodel} defines $\rho_v\in(0,1]$; values above one are allowed here to emulate a wider spread of effective compute capacity, since $\rho_v N_v$ enters the model only as a product. Right: link capacities $c_e$ by link group, in Gbps; server uplinks are identified as links incident to degree-one nodes.}
\label{tab:substrate}
\centering

\renewcommand{\arraystretch}{1.12}
\begin{minipage}[t]{0.56\textwidth}
\centering
\begin{tabular}{lll}
\toprule
\textbf{Parameter} & \textbf{Value} & \textbf{Meaning} \\
\midrule
CPU model & AMD EPYC 7763 & reference host (cycle counts) \\
$f_v$ (GHz) & $\mathcal{U}(3.5,5.5)$ & endpoint clock frequency \\
$N_v$ & $\mathcal{U}_{\mathbb{Z}}(256,1024)$ & logical CPUs \\
$\eta_v$ & $\mathcal{U}(0.75,0.99)$ & CPU efficiency \\
$\rho_v$ & $\mathcal{U}(0.5,1.5)$ & \IP{compute multiplier} \\
$\alpha_v$ & $\mathcal{U}(0.05,0.2)$ & non-cryptographic overhead \\
$\delta_v$ (s) & $0$ & additive I/O delay \\
\bottomrule
\end{tabular}
\end{minipage}\hfill
\begin{minipage}[t]{0.42\textwidth}
\centering
\begin{tabular}{ll}
\toprule
\textbf{Link group} & \textbf{Capacity $c_e$ (Gbps)} \\
\midrule
Fat-tree server--edge & $\mathcal{U}_{\mathbb{Z}}(90,120)$ \\
Fat-tree edge--aggregation & $\mathcal{U}_{\mathbb{Z}}(50,80)$ \\
Fat-tree aggregation--core & $\mathcal{U}_{\mathbb{Z}}(90,120)$ \\
Jellyfish switch--switch & $\mathcal{U}_{\mathbb{Z}}(80,140)$ \\
Jellyfish server uplink & $\mathcal{U}_{\mathbb{Z}}(500,800)$ \\
\bottomrule
\end{tabular}
\end{minipage}
\end{table*}

\subsection{Cryptographic traffic classes}
\label{sec:traffic_classes}

Both experiments use two traffic classes, $X\in\{\mathrm L,\mathrm H\}$. The classes share the same endpoint-selection rules and authentication and symmetric primitives, but differ in the key-establishment mechanism $\mathrm{KEM}_i$ of the profiles $\kappa_i$ and in the distribution of the requested handshake rate $\lambda_i$. The class-specific rate ranges and the rationale for their choice are given below.

The light class is ML-KEM based:
\begin{equation}
{
\begin{aligned}
   \kappa^{\mathrm{L}}_1
&=(\text{ML-DSA-87},\text{ML-KEM-1024},\\
&\qquad\text{AES-256-GCM},5),\\
\kappa^{\mathrm{L}}_2
&=(\text{ML-DSA-65},\text{ML-KEM-768},\\
&\qquad\text{AES-256-GCM},3).
\end{aligned}
}
    \label{eq:light_class_profiles}
\end{equation}
The heavy class keeps the same signature and symmetric primitives but replaces ML-KEM with Classic McEliece:
\begin{equation}
{
\begin{aligned}
\kappa^{\mathrm{H}}_1
&=(\text{ML-DSA-87},\text{McEliece-8192128pcf},\\
&\qquad\text{AES-256-GCM},5),\\
\kappa^{\mathrm{H}}_2
&=(\text{ML-DSA-65},\text{McEliece-460896},\\
&\qquad\text{AES-256-GCM},3).
\end{aligned}
}
    \label{eq:heavy_class_profiles}
\end{equation}
For $X\in\{\mathrm{L},\mathrm{H}\}$, each connection draws its profile uniformly from the two profiles of its class,
\begin{equation}
    {
    \begin{gathered}
        \kappa_i\in\{\kappa^X_1,\kappa^X_2\},\\
        \Pr[\kappa_i=\kappa^X_1]=\Pr[\kappa_i=\kappa^X_2]=\tfrac12,
    \end{gathered}
    }
    \label{eq:kappa_class_uniform}
\end{equation}
so each class is a small distribution over PQC configurations rather than a single deterministic algorithm choice, with security levels $3$ and $5$ equally represented.

\begin{table*}[t]
\caption{Handshake byte budget $B$ of the four PQC profiles, following the TLS~1.3 decomposition of Ref.~\cite{PQCmodel}: a fixed protocol part $B_{\mathrm{fix}}=700$~B plus the KEM public key $b_{\mathrm{pk}}$, the KEM ciphertext $b_{\mathrm{ct}}$, the certificate $b_{\mathrm{cer}}$ (signature public key and certificate signature), and the handshake signature $b_{\mathrm{sig}}$. The last column is the nominal bandwidth demand $8B\lambda_i$, before endpoint clipping, at the mean requested rate of the class ($65$~kconn/s for the light class, $6.5$~kconn/s for the heavy class).}
\label{tab:byte_budget}
\centering

\renewcommand{\arraystretch}{1.12}
\begin{tabular}{llrrrrrr}
\toprule
\textbf{Class} & \textbf{Profile} & $b_{\mathrm{pk}}$ (B) & $b_{\mathrm{ct}}$ (B) & $b_{\mathrm{cer}}$ (B) & $b_{\mathrm{sig}}$ (B) & $B$ (kB) & \textbf{Demand (Gbps)}\\
\midrule
Light & $\kappa^{\mathrm L}_1$: ML-KEM-1024 / ML-DSA-87 & 1\,568 & 1\,568 & 7\,219 & 4\,627 & 15.7 & 8.2\\
Light & $\kappa^{\mathrm L}_2$: ML-KEM-768 / ML-DSA-65 & 1\,184 & 1\,088 & 5\,261 & 3\,309 & 11.5 & 6.0\\
Heavy & $\kappa^{\mathrm H}_1$: McEliece-8192128pcf / ML-DSA-87 & 1\,357\,824 & \IP{240} & 7\,219 & 4\,627 & 1\,370.6 & 71.3\\
Heavy & $\kappa^{\mathrm H}_2$: McEliece-460896 / ML-DSA-65 & 524\,160 & 156 & 5\,261 & 3\,309 & 533.6 & 27.7\\
\bottomrule
\end{tabular}
\end{table*}

{
The two classes are best understood through their byte budgets $B_i$, which depend on $\kappa_i$ alone and are listed in Table~\ref{tab:byte_budget}. A light handshake transmits about $12$--$16$~kB, dominated by the ML-DSA certificate and signature; a heavy handshake transmits $0.5$--$1.4$~MB, almost entirely the Classic McEliece public key. The heavy class is therefore a large-key stress test: it keeps the authentication primitives fixed and \IP{examines a KEM class whose mean handshake byte budget is about seventy times larger than that of the light class}.

To keep the experiment within the capacity range of the fabrics, the heavy class requests ten times fewer connections per second than the light class, $\lambda_i\sim\mathcal U_{\mathbb Z}(5\,000,8\,000)$~conn/s against $\mathcal U_{\mathbb Z}(50\,000,80\,000)$~conn/s. \IP{Both rate ranges define synthetic handshake stress tests.} Even so, before endpoint clipping a heavy connection requests on average about $50$~Gbps of link capacity against $7$~Gbps for a light connection: the tenfold reduction in $\lambda_i$ is outweighed by the seventyfold increase in $B_i$. Since the fat-tree edge--aggregation links have $c_e=50$--$80$~Gbps and the Jellyfish fabric links $c_e=80$--$140$~Gbps (Table~\ref{tab:substrate}), a single heavy connection is comparable to a link, whereas a link can carry roughly ten light connections.

In addition, Classic McEliece key generation, performed by the client at every handshake, is far more expensive than ML-KEM key generation, so the endpoint-feasible rate $\lambda_i^{\mathrm{ep}}$ of Eq.~(\ref{eq:endpoint_feasible_rate}) may also fall below $\lambda_i$ when several heavy connections share an endpoint. All routing policies receive the same adjusted bandwidth demands. Differences between policies therefore arise from their route selection, bandwidth allocation, and path retention.
}

\subsection{Demand construction}
\label{sec:demand_construction}

The topology determines which nodes are eligible endpoints, but the demand construction is otherwise common. If a regime has one admissible endpoint set $\mathcal{A}$, candidate pairs are sampled from
\begin{equation}
    {\mathcal{Q}}
    =
    \big\{\{u,v\}:u,v\in\mathcal{A},\,u\neq v\big\}.
    \label{eq:pairs_one_set}
\end{equation}
If a regime has two endpoint sets, $\mathcal{A}_1$ and $\mathcal{A}_2$, candidates are sampled from
\begin{equation}
    {\mathcal{Q}}
    =
    \big\{{\{u,v\}}:u\in\mathcal{A}_1,\,
    v\in\mathcal{A}_2,\,u\neq v\big\}.
    \label{eq:pairs_two_sets}
\end{equation}
In both cases self-loops are excluded and unordered physical pairs are unique within one traffic matrix. Thus, the load parameter $M$ counts distinct end-to-end services rather than repeated copies of the same source--destination pair.
Equivalently, if the sampled demands are indexed by $i=1,\ldots,M$ with endpoints $(s_i,t_i)$, then
\begin{equation}
    s_i\neq t_i
    \qquad\text{and}\qquad
    \{s_i,t_i\}\neq\{s_j,t_j\}
    \quad\text{for all } i\neq j .
    \label{eq:unique_pairs}
\end{equation}
Here $\{s_i,t_i\}$ denotes the unordered pair (set) of endpoints, using curly-brace set notation rather than the ordered pair $(s_i,t_i)$; the second condition of Eq.~(\ref{eq:unique_pairs}) therefore treats opposite directions as the same physical endpoint pair for sampling purposes, so a physical pair cannot be sampled twice with source and destination swapped. Any directionality (e.g.\ for north--south traffic) is assigned only after this unique-pair selection step, by choosing uniformly at random which endpoint of the pair is the client $s_i$.

After selecting the endpoint pairs, we assign each connection $i$ a requested TLS-like handshake rate $\lambda_i$, sampled independently from the uniform distribution in Eq.~(\ref{eq:lambda_dist_general}), using the bounds specified for its traffic class in Sec.~\ref{sec:traffic_classes}. Requested rates therefore vary across connections within a fixed class-specific range. Independently, we assign one of the class's two cryptographic profiles $\kappa_i$, each with probability $1/2$, as in Eq.~(\ref{eq:kappa_class_uniform}).

\subsection{Experimental procedure}
\label{sec:procedure}

{
For each network, we test every combination of traffic class, endpoint regime, and load. Each combination is evaluated using $100$ independently sampled traffic matrices, with every matrix tested under all three routing policies. The procedure is detailed below and summarised in Table~\ref{tab:factor_grid}.
\begin{enumerate}
    \item \textbf{Fix the network.} For each topology family, one instance $\mathcal N=(V,E,\mathbf c)$ with node specifications $\{\sigma_v\}$ is generated once, with a recorded random seed, by sampling the graph and then the parameters of Table~\ref{tab:substrate}. The instance is never regenerated: all variability reported below comes from the traffic.
    \item {\textbf{Fix a condition.} Choose a traffic class $X\in\{\mathrm L,\mathrm H\}$, an endpoint regime (east--west or north--south on the fat-tree, outer servers on Jellyfish), and a load $M\in\{50,60,\ldots,150\}$. Together these specify one experimental condition.}
    \item {\textbf{Sample the traffic.} For each condition, draw $R=100$ independent traffic matrices. In repeat $\varrho=1,\ldots,R$, the matrix is $\mathcal D^{(\varrho)}=\{d_i\}_{i=1}^{M}$, with each demand record $d_i=(s_i,t_i,\lambda_i,\kappa_i)$ as in Eq.~(\ref{eq:demand_record}). Sample $M$ unique unordered endpoint pairs from $\mathcal Q$ and assign their client and server roles as in Sec.~\ref{sec:demand_construction}; then sample each requested rate $\lambda_i$ from the class law and each profile $\kappa_i$ from Eq.~(\ref{eq:kappa_class_uniform}). Thus, $M$ counts demand records, whereas $\lambda_i$ specifies the requested handshake rate of record $i$. The matrix specifies the offered traffic; routes have not yet been assigned.}
    \item {\textbf{Convert and route.} For each matrix $\mathcal D^{(\varrho)}$, compute the endpoint-feasible rates $\lambda_i^{\mathrm{ep}}$ and corresponding bandwidth demands $b_i$ using Eqs.~(\ref{eq:endpoint_feasible_rate})--(\ref{eq:demand_conversion}). These quantities are common to all routing policies. Evaluate the same demands under each of the three policies $\pi$, obtaining the allocated bandwidths $a_i$ and link flows $f_{i,e}$. Convert the allocations into satisfied handshake rates $r_i$ using Eq.~(\ref{eq:back_conversion}). Each evaluation therefore supplies the quantities needed to compute the observables of Sec.~\ref{sec:observables}.}
    \item \textbf{Aggregate.} The primary observables are computed per evaluation and summarised by their mean and sample standard deviation over the $R$ repeats. The plotted coverage penalty is a difference of class means, and the plotted routing efficiency is the ratio of mean coverage to mean link consumption. Curves are plotted against $M$; where shown, bands indicate one standard deviation.
\end{enumerate}

\begin{table*}[t]
\caption{Experimental factor grid. Each condition combines traffic class, endpoint regime, and load $M$; $100$ traffic matrices and three routing policies give $300$ compute jobs per condition. Shared settings span both experiment columns.}
\label{tab:factor_grid}
\centering

\renewcommand{\arraystretch}{1.05}
\setlength{\tabcolsep}{7pt}
\begin{tabularx}{0.94\textwidth}{@{}>{\raggedright\arraybackslash}p{0.25\textwidth}>{\raggedright\arraybackslash}X>{\raggedright\arraybackslash}X@{}}
\toprule
\textbf{Factor} & \textbf{Fat-tree} & \textbf{Jellyfish}\\
\midrule
Endpoint regimes & East--west: host--host\newline North--south: host--core & Outer server--outer server\\
\midrule
Traffic classes & \multicolumn{2}{l@{}}{Light: $\lambda_i\sim\mathcal U_{\mathbb Z}(50\,000,80\,000)$ conn/s}\\
 & \multicolumn{2}{l@{}}{Heavy: $\lambda_i\sim\mathcal U_{\mathbb Z}(5\,000,8\,000)$ conn/s}\\
Offered load $M$ & \multicolumn{2}{l@{}}{$50,60,\ldots,150$ connections ($11$ values)}\\
Client role & \multicolumn{2}{l@{}}{Uniformly chosen from the sampled endpoint pair}\\
Repeats & \multicolumn{2}{l@{}}{$100$ independent matrices per condition}\\
Routing policies & \multicolumn{2}{l@{}}{OSPF-like; fractional $1$-path; fractional $5$-path}\\
\midrule
Conditions & $2\times2\times11=44$ & $2\times1\times11=22$\\
Traffic matrices & $4\,400$ & $2\,200$\\
\textbf{Compute jobs} & $44\times100\times3=\mathbf{13\,200}$ & $22\times100\times3=\mathbf{6\,600}$\\
\bottomrule
\end{tabularx}
\end{table*}

{
For a fixed network instance, experimental condition, and routing policy, let $O^{(\varrho)}$ denote an observable of Sec.~\ref{sec:observables} measured in repeat $\varrho$. Its reported mean is
\begin{equation}
    \langle O\rangle
    =
    \frac{1}{R}\sum_{\varrho=1}^{R}O^{(\varrho)},
    \qquad R=100.
    \label{eq:experiment_average}
\end{equation}
The same traffic matrix is reused across policies within each repeat, so policy comparisons use identical endpoint pairs, requested rates, and cryptographic profiles.
}

The experiments comprise $13\,200$ fat-tree and $6\,600$ Jellyfish compute jobs. The instance sizes keep the LP evaluations tractable, while the load sweep increases the number of connections competing for endpoint and link resources.
}

\subsection{Observables}
\label{sec:observables}

A \textbf{route} is the network allocation of a single connection record $d_i$; depending on the routing policy it consists of one path component (OSPF-like routing or fractional $1$-path routing) or up to five path components (fractional $5$-path routing). For each connection $i\in\mathcal I$ the evaluation returns the quantities of Sec.~\ref{sec:demand_conversion}: the original requested rate $\lambda_i$, the endpoint-adjusted bandwidth demand $b_i$, the allocated bandwidth $a_i$, the satisfied rate $r_i$, the service ratio $q_i=r_i/\lambda_i$, and the link flows $f_{i,e}$. Let
\begin{equation}
    F_e=\sum_i f_{i,e}
    \label{eq:edge_aggregate_flow}
\end{equation}
be the aggregate flow on link $e$. The following observables describe service delivery, network resource use, and security composition.

\paragraph{Demand coverage.}
The primary benchmark quantity is demand coverage,
\begin{equation}
    C_{\mathrm{dem}}
    =
    \frac{\sum_i r_i}{\sum_i \lambda_i}.
    \label{eq:demand_coverage}
\end{equation}
This lies in $[0,1]$; a value of $1$ means all requested demand is satisfied. The denominator uses the original requested rates $\lambda_i$, before endpoint adjustment. Coverage therefore includes both the reduction imposed by endpoint limits and any further shortfall after routing. By contrast, $a_i/b_i$ measures how fully routing serves connection $i$'s adjusted bandwidth demand. Because $q_i=r_i/\lambda_i$, demand coverage is the demand-weighted counterpart of average QoS:
\begin{equation}
    C_{\mathrm{dem}}
    =
    \frac{\sum_i \lambda_i q_i}{\sum_i \lambda_i}.
    \label{eq:demand_coverage_weighted}
\end{equation}
It coincides with the unweighted average QoS $\bar q=M^{-1}\sum_i q_i$ when all requested demands $\lambda_i$ are equal; with the mild rate heterogeneity used here the two differ by less than $0.003$ in every run, so we report $C_{\mathrm{dem}}$ rather than $\bar q$. In all $19\,800$ evaluations, every endpoint pair was connected by a path with positive link capacities. Thus, disconnection does not contribute to the shortfall $1-C_{\mathrm{dem}}$; the shortfall can reflect endpoint limits, link-capacity constraints under the selected routing policy, and allocation discarded by the path-retention heuristic.

We also record the total served handshake rate,
{\color{black}
\begin{equation}
    r_{\mathrm{tot}}
    =\sum_i r_i
    =C_{\mathrm{dem}}\sum_i\lambda_i,
    \label{eq:total_served_rate}
\end{equation}
}
measured in handshakes per second. Coverage describes the fraction of requested demand that is served, whereas $r_{\mathrm{tot}}$ describes the absolute rate delivered. We use its mean $\langle r_{\mathrm{tot}}\rangle$ over the $R=100$ traffic matrices, calculated as in Eq.~(\ref{eq:experiment_average}), to compare how the served rate changes with load.

\IP{For a fixed traffic matrix, demand coverage is proportional to $\sum_i a_i/B_i$. Thus, the bandwidth objective of Eq.~(\ref{eq:mcf_maxthroughput}) need not maximise handshake coverage when the byte budgets differ.}

\paragraph{Coverage penalty.}
The cost of the heavy class relative to the light class at the same load, endpoint regime, and routing policy is
\begin{equation}
{
    \Delta C_{\mathrm{dem}}
    =
    \langle C_{\mathrm{dem}}^{\mathrm H}\rangle-\langle C_{\mathrm{dem}}^{\mathrm L}\rangle,
}
    \label{eq:coverage_penalty}
\end{equation}
where the brackets denote averages over repeats. Negative values indicate lower coverage for the heavy class, as observed throughout the tested conditions; the sign is an empirical result. A value closer to zero indicates a smaller coverage gap between the classes.

\paragraph{Network stress.}
Network congestion averages the utilisation of all links after the admitted flows have been merged,
\begin{equation}
    \bar\gamma
    =
    \min\!\left(
        1,
        \frac{1}{|E|}\sum_{e\in E}\frac{F_e}{c_e}
    \right),
    \label{eq:network_congestion}
\end{equation}
and link consumption normalises the total flow--link usage by the total capacity of the fabric,
\begin{equation}
    \Phi
    =
    \frac{\sum_{e\in E} F_e}{\sum_{e\in E} c_e}.
    \label{eq:link_consumption}
\end{equation}
For feasible allocations, $\bar\gamma$ weights every link equally, whereas $\Phi$ is the capacity-weighted mean link utilisation. The two coincide when all link capacities are equal. For path-based allocations, $\Phi$ also equals the total allocated bandwidth multiplied by the allocated-bandwidth-weighted mean path length, divided by the total link capacity. Both quantities measure the resources used by admitted traffic; their values depend on both the admitted volume and its routes, and low averages can coexist with individual saturated links.

\paragraph{Routing efficiency.}
\textit{Routing efficiency} measures demand coverage per unit of link resource:
\begin{equation}
    \xi_r = \frac{C_{\mathrm{dem}}}{\Phi}.
    \label{eq:routing_efficiency}
\end{equation}
Larger values indicate more coverage per unit of normalised link consumption. Coverage refers to the original requests, so this ratio measures overall service per unit of link resource, including the effects of endpoint limits. For a fixed traffic mix and similar path lengths, $\xi_r$ decreases approximately as $1/M$ because the total offered demand grows with $M$. Comparisons between policies at fixed load are informative, but comparisons between instances also depend on the total capacity used to normalise $\Phi$. The figures report $\langle C_{\mathrm{dem}}\rangle/\langle\Phi\rangle$, the ratio of the two means over repeats, rather than the mean of the per-evaluation ratio.

\paragraph{Security composition.}
Each connection carries a security level $\ell_i\in\{3,5\}$. We report the unweighted mean level, the demand-weighted level, and the bandwidth-weighted level,
\begin{equation}
{
    \bar{\ell}=\frac{1}{M}\sum_i\ell_i,
    \qquad
    \bar{\ell}_{\mathrm{dem}}
    =
    \frac{\sum_i \lambda_i\ell_i}{\sum_i \lambda_i},
    \qquad
    \bar{\ell}_{\mathrm{bw}}
    =
    \frac{\sum_i a_i\ell_i}{\sum_i a_i}.
}
    \label{eq:security_levels}
\end{equation}
The first two depend only on the traffic matrix and are therefore identical for the three routing policies; their expected value is $4$ by Eq.~(\ref{eq:kappa_class_uniform}). The third weights each level by the bandwidth actually allocated, which depends on both endpoint adjustment and routing. It is not expected to equal $\bar\ell_{\mathrm{dem}}$ even at full coverage, because a level-$5$ handshake carries more bytes than a level-$3$ handshake (Table~\ref{tab:byte_budget}): when all requested handshakes are served ($a_i=b_i$ and $\lambda_i^{\mathrm{ep}}=\lambda_i$ for all $i$), Eq.~(\ref{eq:demand_conversion}) gives $\bar\ell_{\mathrm{bw}}=\sum_iB_i\lambda_i\ell_i/\sum_iB_i\lambda_i$, which is $4.15$ for the light class and about $4.4$ for the heavy class under the sampling laws used here. Both endpoint adjustment and routing can change this balance by reducing service differently across security levels. Endpoint adjustment is common to all policies for a given traffic matrix, so differences between policies arise from routing. \IP{The bandwidth objective of the fractional policies can also influence this admitted composition.} The figures of Sec.~\ref{sec:results} show $\bar\ell$ and $\bar\ell_{\mathrm{bw}}$.

\section{Results}
\label{sec:results}

We examine how much of the requested handshake traffic each network can serve as the number of connections $M$ increases, and how much link capacity it uses to do so. We compare the light and heavy cryptographic traffic classes and assess the benefit of coordinated routing on the fat-tree and Jellyfish. The analysis covers traffic between hosts in the fat-tree, between hosts and core switches in the fat-tree, and between the selected outer servers in Jellyfish.

Demand can remain unserved for three reasons. The endpoints may lack the processing capacity to perform all requested handshakes. The network may have insufficient link capacity to carry the resulting traffic, even with the best possible routing. A routing policy may also leave usable capacity unexploited because of how it selects paths and allocates bandwidth. For fractional routing, some allocated bandwidth can additionally be lost when only one or five paths are retained.

The model first reduces rates where endpoint processing is insufficient, a step called endpoint clipping. It then routes the resulting bandwidth demands subject to the link capacities. Equation~(\ref{eq:back_conversion}) expresses the service delivered after these two stages. All policies receive the same demands after endpoint clipping and operate on the same network. Differences in coverage between policies therefore show how their route selection, bandwidth allocation, and path retention affect the use of the available capacity.

We begin with demand coverage, the fraction of requested handshakes that is served. We then examine link utilisation, the coverage difference between the two traffic classes, and the link capacity used per unit of coverage. Finally, we consider the security levels represented in the served traffic and bring the findings together in Sec.~\ref{sec:results_comparison}. All curves summarise the $100$ traffic matrices for each experimental condition, as described in Sec.~\ref{sec:procedure}.

\subsection{Demand coverage}
\label{sec:results_coverage}

\begin{figure*}[t]
\centering
\includegraphics[width=\textwidth]{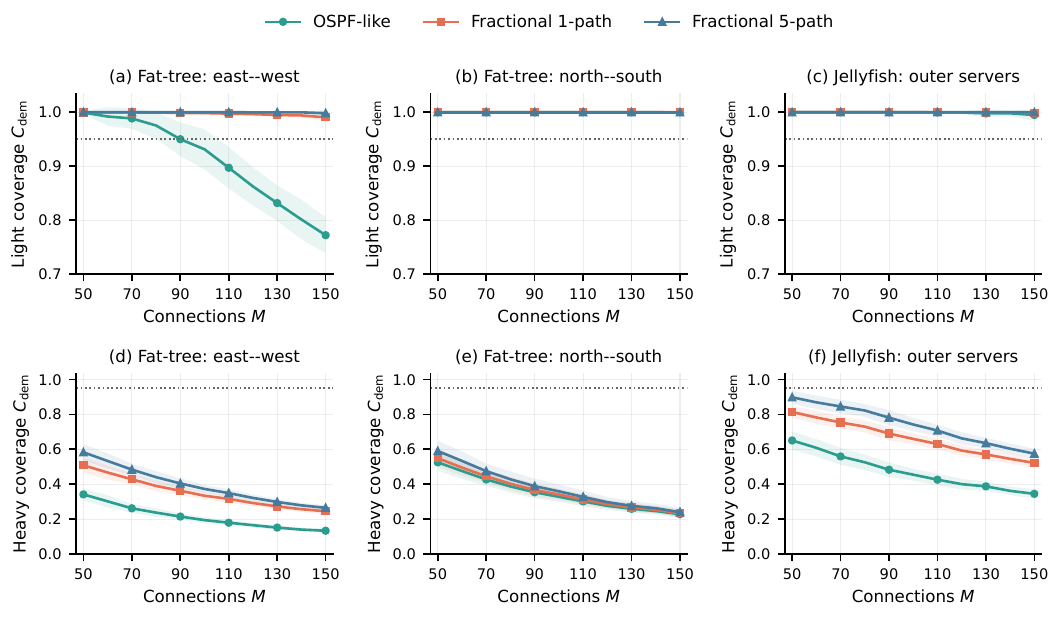}
\caption{Demand coverage versus the number of connections $M$. Columns show fat-tree east--west, fat-tree north--south, and Jellyfish outer-server traffic; rows show the light and heavy classes. Lines are means over $100$ traffic matrices; bands show one standard deviation across these matrices, i.e.\ the spread of outcomes between sampled traffic patterns, not the uncertainty of the mean. The dotted line marks the $0.95$ coverage reference. All panels within a row share the same vertical scale; the light-class scale is expanded to make small losses visible.}
\label{fig:results_coverage}
\end{figure*}

Figure~\ref{fig:results_coverage} compares demand coverage for both traffic classes and all three endpoint regimes. Coverage is measured against the rates originally requested, so it includes losses caused by endpoint processing limits as well as routing. The clearest difference is between the traffic classes. Before endpoint clipping, a heavy-class connection requests about seven times the bandwidth of a light-class connection, despite requesting ten times fewer handshakes per second (Table~\ref{tab:byte_budget}).

The light class achieves almost complete coverage. The only appreciable loss occurs for fat-tree east--west traffic under OSPF-like routing. Its coverage falls below $0.95$ at $M=100$, and more than a fifth of the demand is unserved at $M=150$. In the same regime, both fractional policies remain above $0.99$. All policies serve essentially all light-class demand for fat-tree north--south traffic and on Jellyfish.

\begin{figure*}[t]
\centering
\includegraphics[width=\textwidth]{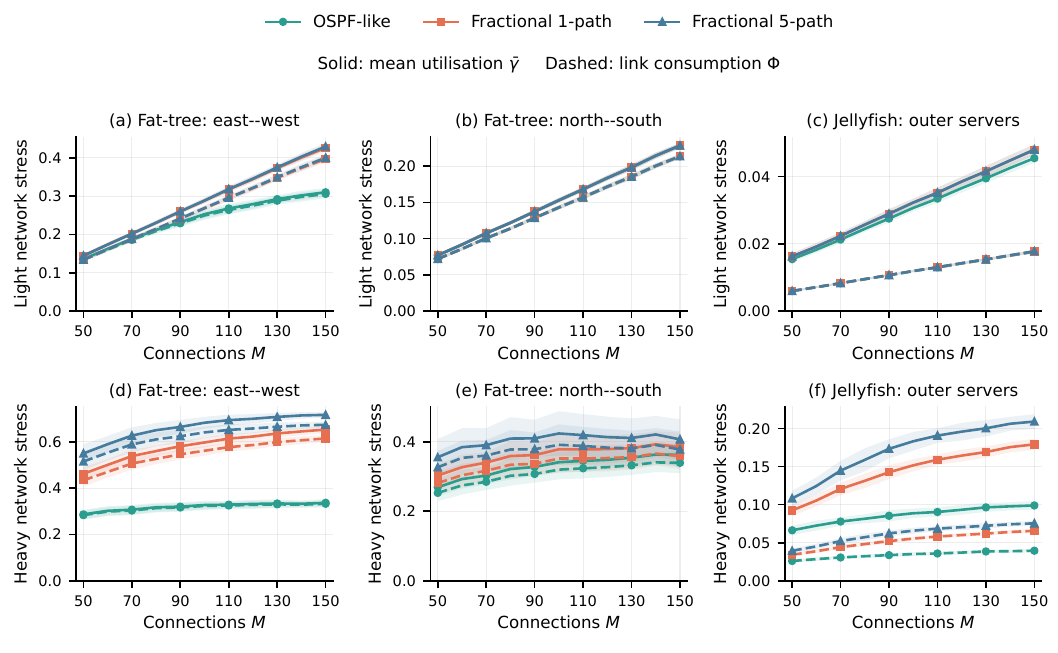}
\caption{Network stress for the same classes and endpoint regimes as Fig.~\ref{fig:results_coverage}. Solid lines show mean link utilisation $\bar\gamma$; dashed lines show capacity-normalised link consumption $\Phi$, on the same axis. Bands show one standard deviation over $100$ traffic matrices, as in Fig.~\ref{fig:results_coverage}. Vertical scales differ between panels to resolve the lower Jellyfish values. Low average stress can coexist with substantial demand loss.}
\label{fig:results_stress}
\end{figure*}

The fractional policies show that the fat-tree can serve almost all of the light-class demand even at the highest load. The loss under OSPF-like routing therefore comes from the policy's use of the network. OSPF-like routing chooses paths independently and concentrates traffic on preferred links, whereas the LP coordinates routes and bandwidth allocations across connections. As load increases, the fractional policies continue to use more link capacity, while utilisation under OSPF-like routing begins to level off (Fig.~\ref{fig:results_stress}). Fractional $1$-path routing already recovers almost all of the lost demand, showing that coordination is effective even when each connection uses only one retained path.

The heavy class is below the $0.95$ coverage reference for every policy and endpoint regime already at $M=50$. The tested range therefore shows how its coverage deteriorates as load increases, rather than where losses begin. As $M$ triples, coverage decreases by roughly $50$ to $60\%$ on the fat-tree and by $35$ to $50\%$ on Jellyfish. Besides requiring more bandwidth, these profiles require costly Classic McEliece key generation at the client for every handshake (Sec.~\ref{sec:traffic_classes}). Both link capacity and endpoint processing can therefore limit service.

Coordinated routing substantially improves heavy-class coverage, but the gain differs between the networks. For fat-tree east--west traffic, fractional $5$-path routing serves $1.7$ to $2$ times as much demand as OSPF-like routing, with a gain of up to $24$ percentage points. On Jellyfish, the gain reaches $30$ percentage points near $M=90$. Fractional $1$-path routing captures most of the improvement in both cases, and retaining additional paths provides a further increase. The Jellyfish gain remains between $23$ and $30$ percentage points throughout the tested range, whereas the fat-tree east--west gain decreases to about $13$ points at $M=150$.

For heavy-class north--south traffic, the three policies remain within about $6$ percentage points of each other. Each host--core pair has only one three-hop shortest path (Sec.~\ref{sec:topologies}), so OSPF-like routing has no choice among equally short routes. Fractional routing can also use longer paths, for example through another edge switch in the same pod. Its link consumption per unit of coverage is about $5$ to $15\%$ higher than that of OSPF-like routing, consistent with some traffic taking these longer routes (Fig.~\ref{fig:results_stress}).

Endpoint processing may also contribute to the smaller policy differences in this regime. Every connection involves one of only nine core switches. Each core switch therefore handles $M/9$ connections on average, three times the average of $2M/54$ connections per host in the east--west regime. This concentration can increase endpoint clipping and leave less demand for any routing policy to serve. The available metrics do not separate this effect from the restrictions imposed by the paths.

Coverage measures the fraction of demand served, while the total served handshake rate $r_{\mathrm{tot}}$ of Eq.~(\ref{eq:total_served_rate}) measures how many handshakes the network serves per second. We use the mean of the recorded total served rates over the $100$ traffic matrices to examine how the delivered rate changes as $M$ increases. These mean rates are not plotted separately; their percentage changes are reported in the following discussion.

As $M$ increases from $50$ to $150$, the mean total served rate for heavy traffic increases by approximately $17\%$ to $44\%$ on the fat-tree, depending on the endpoint regime and routing policy. Under fractional $5$-path routing, the increase is about $36\%$ for east--west traffic and $22\%$ for north--south traffic. Tripling the number of connections therefore produces a much smaller increase in the handshake rate delivered. For east--west traffic at $M=150$, fractional $5$-path routing serves about twice as many handshakes per second as OSPF-like routing. Since endpoint clipping is identical for both policies, this difference comes from routing and bandwidth allocation. The corresponding mean link utilisations are about $0.7$ and $0.3$, respectively (Fig.~\ref{fig:results_stress}).

On Jellyfish, the mean total served rate for heavy traffic under fractional $5$-path routing increases by about $92\%$ over the same load range, nearly doubling. The tested Jellyfish deployment therefore continues to accommodate substantially more traffic as demand grows, whereas the fat-tree shows more limited growth.

Comparing the requested bandwidth with the capacity near the endpoints helps explain where congestion may arise. For fat-tree east--west traffic at $M=50$, the nominal heavy-class demand associated with the three hosts of an edge switch exceeds the combined capacity of its three uplinks by about $40\%$ on average. Here, nominal demand means the bandwidth requested before endpoint clipping. The links from edge to aggregation switches can therefore limit coverage from the lowest tested load. On Jellyfish, the server uplinks have ample capacity, but the $85$ eligible servers attach to only $16$ switches. At $M=150$, the nominal heavy-class demand per endpoint switch exceeds the combined capacity of its six fabric ports by about $40\%$.

These estimates identify plausible bottlenecks near the endpoints in both networks. They assume that nominal demand is spread evenly across endpoints and omit traffic that does not cross the links being considered, such as traffic between hosts on the same edge switch. They therefore indicate where link capacity may be insufficient, without quantifying the resulting coverage loss.

\subsection{Network stress}
\label{sec:results_stress}

Figure~\ref{fig:results_stress} shows how heavily the admitted traffic uses the network. Mean utilisation $\bar\gamma$ weights each link equally, while link consumption $\Phi$ weights links by their capacity. Both depend on the amount of traffic served and the routes it takes. They must therefore be read alongside coverage: a policy that serves less traffic may also have lower utilisation.

For the light class, both measures grow almost linearly with $M$, as expected when nearly all demand is served. Fat-tree north--south traffic produces about half the stress of east--west traffic because paths between hosts and core switches are shorter than paths between hosts. The main exception to linear growth is OSPF-like east--west routing, whose stress begins to level off as more demand remains unserved.

\begin{figure*}[t]
\centering
\includegraphics[width=\textwidth]{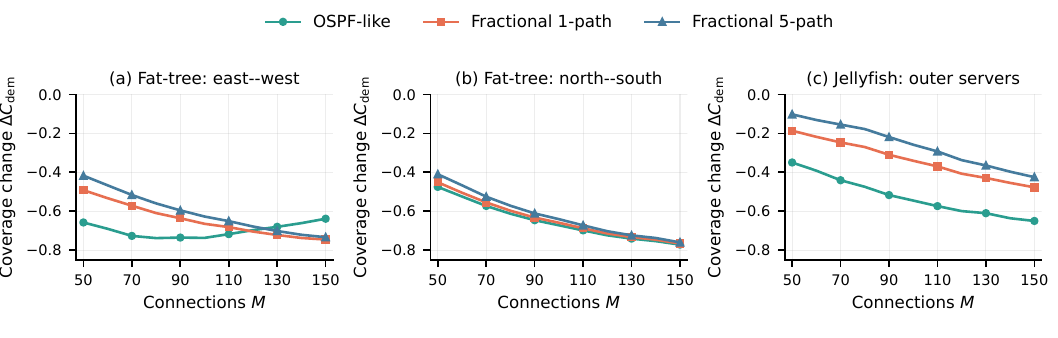}
\caption{Heavy-minus-light coverage difference for the three endpoint regimes. Each curve is the difference between the two class means, with no uncertainty band. All panels use the same vertical scale. A less negative value denotes a smaller difference between classes, which need not imply higher heavy-class demand coverage when the light-class baseline also changes.}
\label{fig:results_penalty}
\end{figure*}

For the heavy class, policies that serve more demand also use more of the network. At the highest load, fractional $5$-path routing for fat-tree east--west traffic has about twice the mean utilisation of OSPF-like routing and serves about twice the demand. OSPF-like utilisation stays close to $0.3$ throughout the tested range. This is consistent with congestion on its preferred paths while other links remain lightly used. For north--south traffic, the policies achieve similar coverage but incur different link consumption, consistent with fractional routing using some longer paths.

On Jellyfish, the separation between policies is much larger for heavy traffic than for light traffic. For the light class, all policies serve almost the same demand and have nearly identical link consumption. The heavy class requests much more bandwidth per connection, making competition for shared fabric links more restrictive. OSPF-like routing keeps its independently selected paths and reduces allocations when they compete for capacity. The fractional policies coordinate both paths and allocations, allowing more demand to be carried through the fabric. Their higher stress therefore accompanies the substantial coverage gains seen in Fig.~\ref{fig:results_coverage}. Endpoint clipping is common to the policies and cannot explain this separation.

The relative increase in Jellyfish stress is also larger than the increase in coverage. At $M=150$, fractional $5$-path routing serves about $67\%$ more heavy-class handshakes per second than OSPF-like routing, while link consumption is about $91\%$ higher and mean utilisation is approximately twice as large. Most of the increase in link consumption accompanies the larger served rate, but link consumption per served handshake is also about $15\%$ higher. Coverage measures the fraction of requested handshakes served, whereas link consumption counts the bandwidth they occupy on every link along their routes. The two measures therefore need not increase in the same proportion.

This additional link consumption is consistent with fractional routing using some longer paths and serving a larger bandwidth share of level-$5$ traffic, whose handshakes carry more bytes (Sec.~\ref{sec:results_security}). The aggregate measurements do not separate these contributions. Mean utilisation also depends on which links carry the traffic: the same bandwidth produces greater utilisation on a lower-capacity link. A larger relative rise in $\bar\gamma$ than in $\Phi$ therefore indicates a change in how traffic is distributed across links of different capacities. Together, these effects explain why the policy differences in stress can be more pronounced than those in coverage.

On Jellyfish, mean utilisation reaches only about $0.21$ and link consumption stays below $0.1$, even when a substantial share of heavy-class demand is unserved. These low averages partly reflect the $360$ server uplinks. Their large capacities increase the denominator of $\Phi$, and the many lightly used uplinks also lower the mean utilisation across all links. Congestion can still occur on individual links between switches, including those attached to the $16$ switches serving the eligible endpoints. Low average stress therefore does not show that the remaining capacity is accessible to the connections that need it. Endpoint clipping can also contribute to the combination of low utilisation and unserved demand.

\subsection{Coverage penalty of the heavy class}
\label{sec:results_penalty}

Figure~\ref{fig:results_penalty} shows the heavy-class coverage minus the light-class coverage, $\Delta C_{\mathrm{dem}}=C_{\mathrm{dem}}^{\mathrm H}-C_{\mathrm{dem}}^{\mathrm L}$, at the same $M$, endpoint regime, and routing policy. A more negative value means a larger coverage disadvantage for the heavy class. The comparison uses each class's own requested handshake rates, so it describes the difference between the complete traffic classes, including their bandwidth and processing requirements.

The coverage gap generally widens as load increases. Where light-class coverage stays close to one, an improvement in heavy-class coverage directly reduces the penalty. This is the case on Jellyfish, where fractional $5$-path routing keeps the penalty $20$ to $30$ percentage points smaller than OSPF-like routing throughout the tested range. For fat-tree north--south traffic, the policies have similar penalties because they achieve similar coverage for both classes.

The penalty must be interpreted with care when light-class coverage also changes. For fat-tree east--west traffic, fractional $5$-path routing has the smaller penalty at low load, but at the highest load its penalty is larger than that of OSPF-like routing, even though it serves almost twice as much heavy-class traffic. The reversal occurs because OSPF-like routing also loses a substantial part of the light-class demand, which narrows the gap between the two classes. Read alongside the demand coverage of each class in Fig.~\ref{fig:results_coverage}, the penalty shows how the disadvantage of heavy traffic depends on routing and load. On Jellyfish, coordinated routing reduces this disadvantage while preserving almost complete light-class coverage. For fat-tree east--west traffic, its benefit is to maintain high light-class coverage while also serving more heavy traffic, even when the gap between the classes remains large.

\begin{figure*}[t]
\centering
\includegraphics[width=\textwidth]{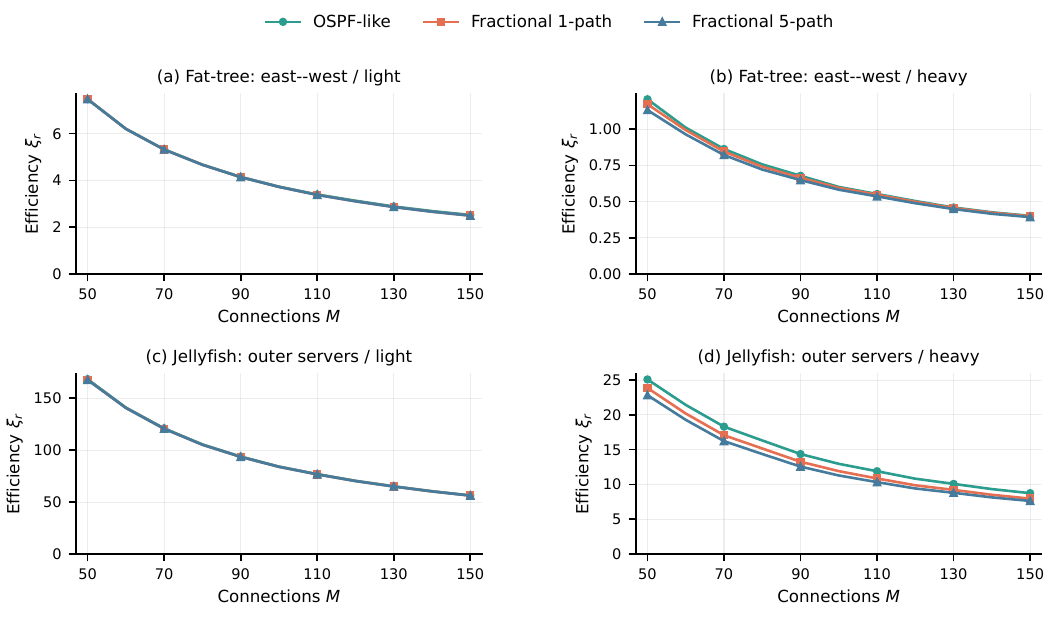}
\caption{Routing efficiency for fat-tree east--west traffic (top) and Jellyfish outer-server traffic (bottom), with light and heavy classes in the left and right columns. Curves show the ratio of the mean coverage to the mean link consumption, $\langle C_{\mathrm{dem}}\rangle/\langle\Phi\rangle$, rather than the mean of per-run ratios. Vertical scales differ. The metric depends on the total capacity of the instance, so absolute values do not provide a capacity-independent ranking of the topologies.}
\label{fig:results_efficiency}
\end{figure*}

\subsection{Routing efficiency}
\label{sec:results_efficiency}

Figure~\ref{fig:results_efficiency} compares the demand coverage obtained per unit of normalised link consumption for fat-tree east--west and Jellyfish traffic. At a fixed load, a larger value means that the policy uses less link capacity for the same coverage. If the cryptographic profiles and path lengths of the served traffic remain similar, reducing its rate lowers both coverage and link consumption in approximately the same proportion. This explains why the light-class policies have nearly identical efficiency even when their coverages differ.

Efficiency decreases approximately as $1/M$ in all cases because coverage is divided by the total requested rate, which grows with $M$. For a fixed average link cost per served handshake, this introduces an inverse dependence on $M$ into the ratio of coverage to link consumption. The decline therefore does not by itself mean that routes become less economical at higher load.

\begin{figure*}[t]
\centering
\includegraphics[width=\textwidth]{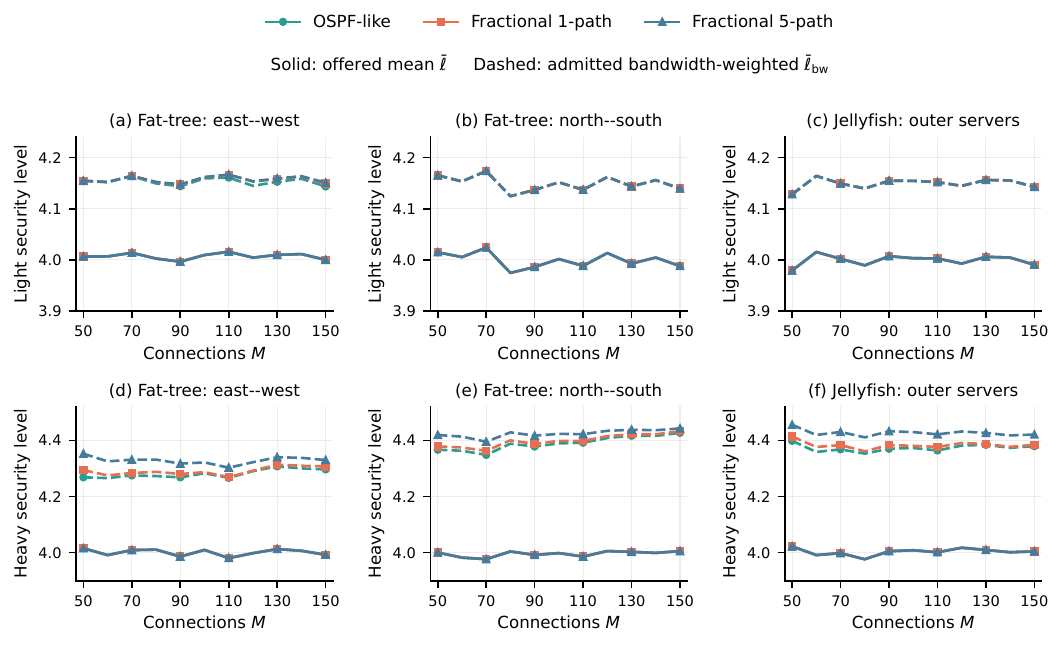}
\caption{Security-level composition for both classes and all endpoint regimes. Solid lines show the mean level of the offered connections, $\bar\ell$; dashed lines show the bandwidth-weighted level of admitted traffic, $\bar\ell_{\mathrm{bw}}$. Points are means over $100$ traffic matrices; no uncertainty bands are shown. Policy differences are small for the light class and more visible for the heavy class, particularly for fat-tree east--west traffic.}
\label{fig:results_security}
\end{figure*}

The heavy class is about six to seven times less efficient than the light class on both networks, consistent with its roughly sevenfold larger nominal bandwidth demand per connection. Within the heavy class, fractional routing serves substantially more traffic than OSPF-like routing but uses slightly more link capacity per unit of coverage. Two effects may contribute. Fractional routing can use longer paths to reach additional capacity, as suggested by the north--south results. It also serves a larger bandwidth share of level-$5$ traffic, whose handshakes carry more bytes (Sec.~\ref{sec:results_security}). The reported efficiency does not separate these effects.

Jellyfish's efficiency is more than an order of magnitude higher than the fat-tree's. Much of this difference comes from the definition of link consumption: $\Phi$ divides the traffic carried across links by the network's total link capacity, which is about $17$ times larger on Jellyfish. Absolute efficiency values therefore depend strongly on network provisioning. The gains between policies on each network provide a more useful measure of the benefit of coordinated routing.

\subsection{Security composition}
\label{sec:results_security}

Figure~\ref{fig:results_security} compares the security levels of the requested connections with those represented in the traffic actually served. The unweighted mean $\bar\ell$ describes the requested connections and is identical across policies because they receive the same traffic matrix. The bandwidth-weighted mean $\bar\ell_{\mathrm{bw}}$ gives more weight to profiles that receive more bandwidth. It therefore shows whether the share of bandwidth assigned to level-$5$ traffic changes between policies. Endpoint clipping can also affect this share, but it is common to all policies.

For the light class, $\bar\ell_{\mathrm{bw}}$ stays close to the full-service reference of about $4.15$ for all policies. The value exceeds the offered mean of about $4$ because level-$5$ handshakes carry more bytes than level-$3$ handshakes. The admitted bandwidth therefore approximately preserves the mix expected when all requests are served. This remains true for fat-tree east--west traffic under OSPF-like routing, even though more than a fifth of the demand is unserved at the highest load.

For the heavy class, fractional $5$-path routing gives the highest bandwidth-weighted mean in every regime, followed by fractional $1$-path and OSPF-like routing. The differences are modest, at most about $0.085$, and are largest for fat-tree east--west traffic. Since endpoint clipping is common to all policies, this ordering comes from routing: policies that serve more heavy-class demand also allocate a larger bandwidth share to level-$5$ traffic. This is consistent with the larger level-$5$ profiles being more affected by capacity constraints. Their larger byte budgets also contribute to the slightly lower efficiency of fractional routing.

The security composition is therefore approximately preserved for the light class and changes modestly between policies for the heavy class. These means describe the mix of assigned NIST levels in the traffic. They do not assign a security level to the network as a whole.

\subsection{Comparison and limitations}
\label{sec:results_comparison}

Two principal results emerge. The first is the marked difference between light and heavy cryptographic traffic. At the tested rates, the light class achieves almost complete coverage except for fat-tree east--west traffic under OSPF-like routing. The heavy class leaves substantial demand unserved on both networks from the lowest load. Its mean handshake byte budget is approximately seventy times larger, so it requests roughly seven times more bandwidth per connection even with a tenfold lower handshake rate. Network provisioning must therefore account for both the cryptographic profile and the requested handshake rate.

The second result concerns the effectiveness of coordinated routing on the two network structures. For the heavy class, the benefit is larger and more sustained on the tested Jellyfish network. Fractional $5$-path routing recovers about $70\%$ of the demand left unserved by OSPF-like routing at $M=50$ and about $35\%$ at $M=150$. On the fat-tree, the recovered share is below $40\%$ for east--west traffic and decreases to about $15\%$ at the highest load. For north--south traffic, it is at most about $15\%$.

The coverage gains, expressed in percentage points, show the same difference. On Jellyfish, the gain stays between $23$ and $30$ percentage points throughout the tested range. For fat-tree east--west traffic, it decreases from $24$ to about $13$ points as the served rate approaches a plateau. Fractional $1$-path routing already captures most of these gains, showing that coordinated route selection and bandwidth allocation can be effective without splitting each connection across several paths. The smaller north--south gain depends more on retaining additional paths.

The path structure helps explain these trends. Jellyfish offers alternative paths through its random regular fabric~\cite{Jellyfish}, which coordinated routing can use to avoid congested links. Fat-tree east--west traffic also has several possible paths, but connections between different edge switches must cross the links from the edge to the aggregation layer. Those links provide less total capacity than the host links feeding them. The shrinking routing gain is consistent with this shared constraint becoming more restrictive as load increases. For host--core traffic, the unique shortest path further limits the choices available to OSPF-like routing. The results thus indicate substantial potential for traffic engineering in the tested Jellyfish network and a need to consider access-layer capacity alongside routing in the fat-tree.

The policy comparisons measure how much service routing can recover. They do not directly measure how much demand is removed by endpoint processing limits, but coverage places an upper bound on that loss. Since $q_i\leq\lambda_i^{\mathrm{ep}}/\lambda_i$, the fraction removed before routing cannot exceed $1-C_{\mathrm{dem}}$ of the best policy. For example, serving about $90\%$ of the requested demand means that endpoint clipping can have removed at most about $10\%$. This is the bound for the heavy class on Jellyfish at $M=50$; the corresponding fat-tree bound is about $40\%$. Where light-class coverage is almost complete, endpoint clipping must be negligible. Demand may also remain unserved because of network constraints or because bandwidth is discarded when the fractional solution is reduced to one or five paths.

Further analysis would test how broadly these findings apply. Each topology family is represented by one fixed network, and the $100$ repeats vary the traffic rather than the topology. The networks also differ in total capacity, size, and the number of connections sharing each endpoint. Comparing policies on the same network establishes the benefit of coordination under those conditions. Comparing the size of that benefit across the two networks still involves differences in provisioning and load as well as topology. Similarly, the light and heavy classes differ in both handshake size and processing cost, so the experiment does not measure the separate contribution of each.

Several additional measurements would help distinguish these contributions. Reporting the rates remaining after endpoint clipping would identify the processing loss directly. Coverage before the LP solution is reduced to a fixed number of paths would reveal the loss caused by that reduction. Path lengths and utilisation on individual links would help locate network bottlenecks. Optimising bandwidth allocation while keeping the OSPF-like paths fixed would separate the benefit of allocation from that of choosing different routes. An equal-cost multi-path baseline and experiments with comparable capacities across several network instances would further test the role of topology. The present results already provide a clear direction for these studies: large-key traffic requires substantially more resources, and coordinated routing can recover considerable demand, with gains that depend on network structure and provisioning.

\section{Conclusion}
\label{sec:conclusion}

We have evaluated how post-quantum handshake traffic uses data-centre networks by combining an endpoint and network throughput model~\cite{PQCmodel} with multiple-unicast routing. The study compares two cryptographic traffic classes and three routing policies on fixed fat-tree and Jellyfish deployments as the number of connections increases. Coverage measures the fraction of requested demand served, while the total served handshake rate shows how much traffic the network delivers per second. Network stress, routing efficiency, and security composition help explain the resources used to provide that service.

The first main result is the difference between light and heavy traffic. Coordinated routing serves almost all ML-KEM/ML-DSA light-class demand at the tested rates. The appreciable loss under OSPF-like routing for fat-tree east--west traffic can therefore be recovered through better use of the same network. The Classic McEliece heavy class requests about seven times more bandwidth per connection despite a tenfold lower handshake rate, and all policies fall below $0.95$ coverage from the lowest tested load. The results demonstrate the importance of considering handshake byte budget and requested rate together, while also accounting for the processing requirements of the selected profiles.

The second main result concerns routing effectiveness across the two network structures. On Jellyfish, fractional $5$-path routing recovers about $70\%$ of the heavy-class demand left unserved by OSPF-like routing at the lowest load and about $35\%$ at the highest. Its coverage gain stays between $23$ and $30$ percentage points. For fat-tree east--west traffic, the recovered share is below $40\%$ and decreases to about $15\%$, while the coverage gain narrows from $24$ to about $13$ points. For north--south traffic, the recovered share is at most about $15\%$. One retained path per connection captures most of the gains for Jellyfish and fat-tree east--west traffic. This shows the benefit of coordinated route selection and bandwidth allocation even without splitting each connection across several paths.

The total served rate clarifies how these differences develop with load. As the number of connections increases from $50$ to $150$, fractional $5$-path routing increases the mean served rate for heavy traffic by about $92\%$ on Jellyfish, compared with $36\%$ for fat-tree east--west traffic and $22\%$ for north--south traffic. Thus, falling coverage can accompany substantial growth in the rate delivered, particularly on Jellyfish. The higher network stress under fractional routing reflects both the additional traffic served and, for heavy Jellyfish traffic, greater link consumption per served handshake. The security composition remains approximately preserved for the light class and changes modestly between policies for the heavy class.

These observations indicate sustained opportunities for traffic engineering on the tested Jellyfish network. On the fat-tree, the more limited growth in served traffic and the nominal bandwidth demands near the hosts point to access-layer provisioning as an important consideration alongside routing. In particular, the estimated heavy-class demand for east--west traffic already exceeds edge-switch uplink capacity at the lowest tested load. This identifies a plausible network constraint, while endpoint processing and the allocation discarded when paths are removed can also contribute to unserved demand.

Each topology family is represented by one fixed deployment, and the deployments differ in capacity, size, and endpoint placement. The results therefore provide directions for provisioning and routing decisions under the tested conditions. Experiments with comparable capacities across several network instances would establish how broadly the differences between the topologies apply. Reporting endpoint-feasible rates, coverage before path retention, and utilisation on individual links would help separate the remaining processing, routing, and capacity constraints. An equal-cost multi-path baseline and hybrid cryptographic profiles are further extensions. The framework provides a practical basis for assessing these choices together before deploying a given mix of post-quantum profiles.


\section*{Acknowledgements.}
S.P. acknowledges support from UKRI via the Integrated Quantum Networks Research Hub (IQN, EP/Z533208/1).

\section*{Data Statement}
The topology generation, traffic generation, and routing evaluation were performed with the digital-twin platform developed by nodeQ Limited, which implements the endpoint and network model of Ref.~\cite{PQCmodel}. The data comprise one fixed topology instance for each network family, $4\,400$ traffic matrices for the fat-tree, $2\,200$ traffic matrices for Jellyfish, and the recorded observables from all $19\,800$ completed routing evaluations. Each traffic matrix was evaluated under all three routing policies. These data and supporting experimental specifications are available from the corresponding author on reasonable request. The evaluation uses proprietary software components.

\FloatBarrier

\end{document}